\pdfoutput=1
\documentclass[preprint,12pt,authoryear]{elsarticle}

\usepackage{amssymb}
\usepackage{amsthm}
\usepackage{amsmath}
\usepackage{bm}
\usepackage{caption}
\usepackage{subcaption}
\usepackage{enumitem}
\usepackage{url}
\usepackage{array,multirow,graphicx}

\usepackage{tikz}
\usepackage{pgfplots}
\pgfplotsset{compat=newest}
\usetikzlibrary{shapes.geometric,arrows,fit,matrix,positioning}
\tikzset
{
    treenode/.style = {circle, draw=black, align=center, minimum size=1cm},
    subtree/.style  = {isosceles triangle, draw=black, align=center, minimum height=0.5cm, minimum width=1cm, shape border rotate=90, anchor=north},
    terminal/.style  = {rectangle, draw=black, align=center, minimum height=0.8cm, minimum width=1cm, anchor=north}
}

\newtheorem{thm}{Theorem} 
 
\newdefinition{rmk}{Remark} 
\newproof{pf}{Proof}
\newproof{pot}{Proof of Theorem \ref{thm2}}
\newproof{pfmew}{Proof of Theorem \ref{thm:mew}}

\newproof{pfmec}{Proof of Theorem \ref{thm:mec}}

\newproof{pfdist}{Proof of Theorem \ref{thm:dist}}

\newproof{pfinert}{Proof of Theorem \ref{thm:inert}}

\newproof{pfrmterms}{Proof of Theorem \ref{thm:rmterms}}

\newproof{pfsobolbart}{Proof of Theorem \ref{thm:sobolbart}}

\newproof{pftied}{Proof of Theorem \ref{thm:tied}}

\newcommand{\x}{\mathbf{x}}
\newcommand{\X}{\mathbf{X}}

\newcommand{\Cov}{\texttt{Cov}}
\newcommand{\Var}{\texttt{Var}}

\newcommand{\1}{\mathbf{1}}
\newcommand{\R}{\mathbf{R}}
\newcommand{\E}{\mathbb{E}}
\newcommand{\Prob}{\mathbb{P}}
\newcommand{\I}{I}

\newcommand{\ens}{\mathcal{E}}

\newcommand{\T}{\mathcal{T}}
\newcommand{\M}{\mathcal{M}}

\newcommand{\iid}{\stackrel{iid}{\sim}}
\makeatletter
\DeclareFontEncoding{LS1}{}{}
\DeclareFontSubstitution{LS1}{stix}{m}{n}
\DeclareMathAlphabet{\mathscr}{LS1}{stixscr}{m}{n}
\makeatother

\newcommand{\dat}{\mathscr{D}}

\journal{Reliability Engineering $\&$ System Safety}

\begin{document}

\begin{frontmatter}

%% Title, authors and addresses

%% use the tnoteref command within \title for footnotes;
%% use the tnotetext command for theassociated footnote;
%% use the fnref command within \author or \address for footnotes;
%% use the fntext command for theassociated footnote;
%% use the corref command within \author for corresponding author footnotes;
%% use the cortext command for theassociated footnote;
%% use the ead command for the email address,
%% and the form \ead[url] for the home page:
%% \title{Title\tnoteref{label1}}
%% \tnotetext[label1]{}
%% \author{Name\corref{cor1}\fnref{label2}}
%% \ead{email address}
%% \ead[url]{home page}
%% \fntext[label2]{}
%% \cortext[cor1]{}
%% \address{Address\fnref{label3}}
%% \fntext[label3]{}

\title{Assessing variable activity for Bayesian regression trees}

%% use optional labels to link authors explicitly to addresses:
%% \author[label1,label2]{}
%% \address[label1]{}
%% \address[label2]{}

% \author{}

\author[1]{Akira Horiguchi\corref{cor1}} \ead{ahoriguchi9991@gmail.com}
\author[2]{Matthew T. Pratola} \ead{mpratola@stat.osu.edu}
\author[3]{Thomas J. Santner} \ead{santner.1@osu.edu}
\cortext[cor1]{Corresponding author}
\address{The Ohio State University, Cockins Hall, 1958 Neil Ave., Columbus, OH 43210, USA}

\begin{abstract}
%% Text of abstract
Bayesian Additive Regression Trees (BART) are non-parametric models that can capture complex exogenous variable effects. In any regression problem, it is often of interest to learn which variables are most active. Variable activity in BART is usually measured by counting the number of times a tree splits for each variable.  Such one-way counts have the advantage of fast computations. Despite their convenience, one-way counts have several issues. They are statistically unjustified, cannot distinguish between main effects and interaction effects, and become inflated when measuring interaction effects. An alternative method well-established in the literature is Sobol\'{} indices, a variance-based global sensitivity analysis technique. However, these indices often require Monte Carlo integration, which can be computationally expensive. This paper provides analytic expressions for Sobol\'{} indices for BART posterior samples. These expressions are easy to interpret and are computationally feasible. Furthermore, we will show a fascinating connection between first-order (main-effects) Sobol\'{} indices and one-way counts. We also introduce a novel ranking method, and use this to demonstrate that the proposed indices preserve the Sobol\'{}-based rank order of variable importance. Finally, we compare these methods using analytic test functions and the En-ROADS climate impacts simulator.
\end{abstract}

%%Graphical abstract
% \begin{graphicalabstract}
% %\includegraphics{grabs}
% \end{graphicalabstract}

%%Research highlights
% \begin{highlights}
% \item Count heuristic in regression trees is misleading
% \item Sobol\'{} indices are more accurate than count heuristic
% \item Sobol\'{} indices for regression trees can be computed exactly
% \item Global sensitivity analysis can be performed in high-dimensional input spaces
% \end{highlights}

\begin{keyword}
%% keywords here, in the form: keyword \sep keyword

%% PACS codes here, in the form: \PACS code \sep code

%% MSC codes here, in the form: \MSC code \sep code
%% or \MSC[2008] code \sep code (2000 is the default)
Bayesian Additive Regression Trees
\sep Global sensitivity analysis 
\sep Sobol\'{} indices 
\sep Nonparametric
\sep Variable importance
\sep Variable activity
\end{keyword}

\end{frontmatter}
%% \linenumbers

%% main text

%%%%%%%%%%%%%%%%%%%%%%%%%%%%%%%%%%%%%%%%%%%%%%%%%%%%%%%%%%%%%%%%%%%%%%%%%%%%
%%%%%%%%%%%%%%%%%%%%%%%%%%%%%%%%%%%%%%%%%%%%%%%%%%%%%%%%%%%%%%%%%%%%%%%%%%%%%%%%%%%%%%%%%%%%%%%%%%%
\section{Introduction} 
\label{sec:intro}

Bayesian Additive Regression Trees (BART) have become an increasingly popular tool for complex regression problems and as emulators of expensive computer simulations \citep{CGM10,Chipman12,Gramacy16}. BART sidesteps the $O(n^3)$ matrix decompositions required by arguably the most popular statistical regression tool, Gaussian processes (GPs) \citep{Santner18}. These cubic matrix operations pose issues whose severity continues to grow in the era of big data. BART, like GPs, can capture complex exogenous variable effects without having to specify their functional forms. 

\begin{figure}[t]
\centering
\begin{subfigure}[t]{0.48\textwidth}
    \begin{tikzpicture}[->,>=stealth', level/.style={sibling distance = 5cm/#1, level distance = 1.5cm}, scale=0.7, transform shape]
    \node [treenode] {$x_2$ \\ $< 0.7$}
    child {
        node [treenode] {$x_1$ \\ $< 0.2$} 
        child { node [terminal] {$\mu_1$} }
        child { node [terminal, gray] {$\mu_2$} }
    }
    child {
        node [treenode] {$x_1$ \\ $< 0.4$}      
        child { node [terminal] {$\mu_3$} }
        child { node [terminal] {$\mu_4$} }
    }
;
\end{tikzpicture}
\caption{Node view of the example tree. 
\newline Input $\x^*~= ~(0.9, 0.6)$ falls into the gray terminal node.}
\label{fig:tree1a}
\end{subfigure}
\begin{subfigure}[t]{0.48\textwidth}
    \centering
    \begin{tikzpicture}[scale=0.7]
        \begin{axis}[
        xmin=0, xmax=1, xlabel={$x_1$}, 
        ymin=0, ymax=1, ylabel={$x_2$}]
            \addplot [dotted, thick] table {
                0 0.7
                1 0.7
            };
            \addplot [dotted, thick] table {
                0.2 0
                0.2 0.7
            };
            \addplot [dotted, thick] table {
                0.4 0.7
                0.4 1
            };
            \addplot [only marks] table {
                0.9 0.6
            };
            \node at (axis cs:0.1,0.35) {$\mu_1$};
            \node at (axis cs:0.6,0.35) [gray] {$\mu_2$};
            \node at (axis cs:0.2,0.85) {$\mu_3$};
            \node at (axis cs:0.7,0.85) {$\mu_4$};
        \end{axis}
    \end{tikzpicture}
    \caption{Level-set view of the example tree.\newline 
    Input 
     $\x^*~= ~(0.9, 0.6)$ is shown as the black point.}
    \label{fig:tree1b}
\end{subfigure}
\caption{Two different views of the same example tree.}
\label{fig:tree1}
\end{figure}
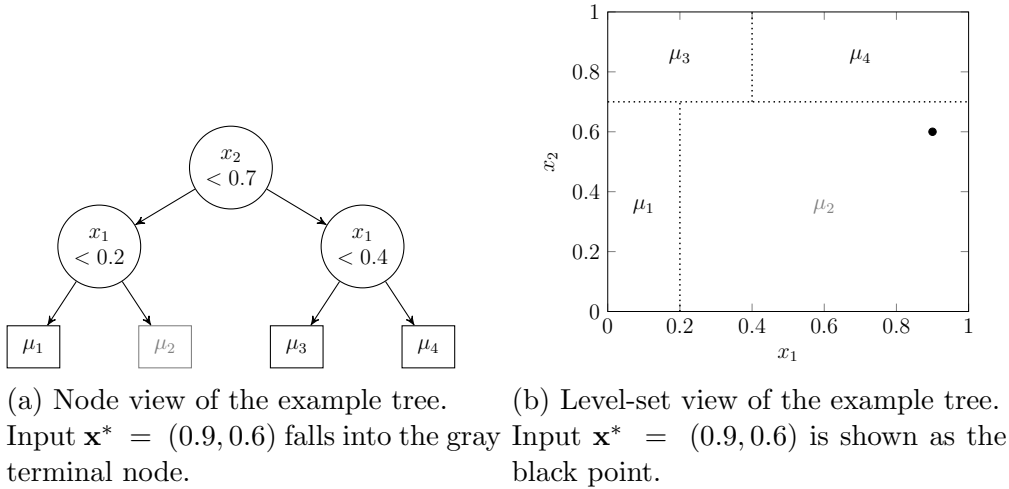

To assess the activity of these exogenous input variables, BART offers a variable count heuristic proposed by \cite{CGM10}, which comes nearly for free once a BART model is fit. This method counts the number of times a variable is included in BART's trees as a split variable. For example, the tree in Figure \ref{fig:tree1a} splits on $x_1$ twice and on $x_2$ once. Using this heuristic, input $x_1$ would be considered to be twice as active as input $x_2$. The idea is that if many nodes in BART's trees split on a variable, then that variable is deemed important in predicting the response. To this day, count-based methods remain the most popular way of assessing input activity in BART. For example, \cite{Bleich14} also rely on these posterior inclusion proportions in their proposed variable selection methods. 

\begin{figure}[t!]
    \centering
    \includegraphics[width=0.4\textwidth]{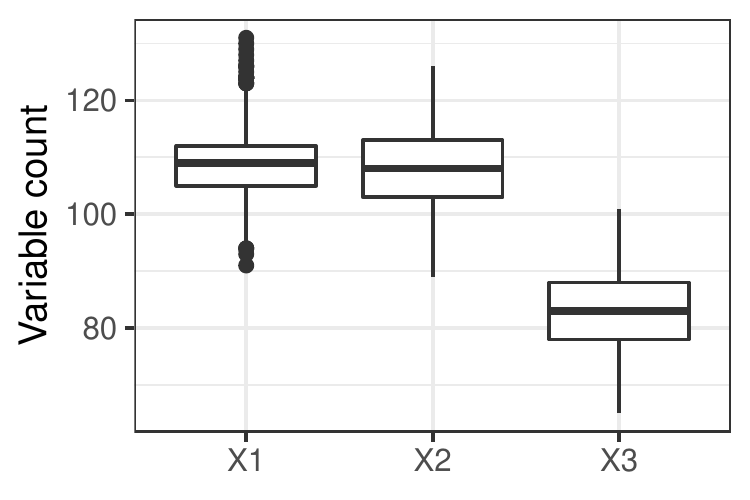}
    \caption{One-way variable counts of 1,000 posterior samples from a BART ensemble trained in data generated from the function $f(\x) = (x_1 - 0.5)(x_2 - 0.5) + 0.5(x_3 - 0.5)$ on the unit hypercube $[0, 1]^3$.}
    \label{fig:count}
\end{figure}

But as \cite{LRW18} note, one-way counts are not theoretically well-understood. Furthermore, their ability to adequately capture even the order of input importance is suspect. Figure \ref{fig:count} shows the variable counts of 1,000 posterior samples from a BART model trained in data generated from the function $f(\x) = (x_1 - 0.5)(x_2 - 0.5) + 0.5(x_3 - 0.5)$ on the unit hypercube $[0, 1]^3$. Marginally, variables $x_1$ and $x_2$ have zero effect on $f(\bm{\cdot})$, which makes variable $x_3$ marginally the most important input. But the variable counts in Figure \ref{fig:count} show $x_1$ and $x_2$ to be more active than $x_3$. Thus, the individual marginal counts seem to conflate the interaction effect between $x_1$ and $x_2$ with their marginal effects.  

To better assess input activity, we may instead use the variance-based global 
sensitivity analysis method introduced by \cite{Sobol93}.
He showed that if $f(\x)$ is a real-valued, square-integrable function on $[0,1]^p$ then $f(\x)$ can be decomposed into a sum
\begin{equation*}
f(\x) = f_0 + \sum_{i=1}^d f_i(x_i) + \sum_{i=1}^d \sum_{j<i} f_{ij}(x_i, x_j) + \cdots + f_{1, 2, \ldots, p} (x_1, x_2, \ldots, x_p),
\end{equation*}
where each summand depends on a subset of $\x$.
Assume that the relative frequency with which the inputs of $f(\x)$ occur can be modeled by $\X = (X_1, X_2, \ldots, X_p)$ where $X_1, \ldots, X_p \iid U(0,1)$. Then if the variance of the $i$th term in the Sobol\'{} expansion which depends on $x_i$ is large, then $x_i$ is deemed important in predicting the response.
Computing these variances and expectations requires Monte Carlo integration when $f(\x)$ is not known in closed form, and hence becomes untractable as the number of inputs increases.

Sobol\'{} indices have been estimated or computed for various types of metamodels. These approaches can be divided into two groups depending on whether they assume the inputs are uncorrelated. 
For uncorrelated inputs, Chapter 7 of \cite{Santner18} provides an overview of GP-based Sobol\'{} indices and their formulae for GPs with certain mean and correlation structures \cite[see also][]{Oakley2004,Chen2005,Chen2006,Marrel2009,Moon2010,Svenson2014}. \cite{Sudret2008} provide analytic expressions for polynomial-chaos-based Sobol\'{} indices, which reduces the computational burden to obtaining the desired polynomial-chaos coefficients. 
Finally, \cite{Gramacy2013} and \cite{Gramacy2010} suggest using Sobol\'{} indices for Dynamic Trees and Treed Gaussian Processes, which use integration approximations via Latin hypercube designs to compute these index estimates.
For possibly correlated inputs, \cite{DaVeiga2009} compute first-order sensitivity indices of local polynomial smoothers and provide theoretical asymptotic properties for the indices. \cite{Wei2015}, motivated by high-dimensional input spaces, show that the random-forest-based permutation variable importance measure converges in $n$ to twice the unnormalized total-effects Sobol\'{} index. The methods from \cite{DaVeiga2009} and \cite{Wei2015} avoid numerical integration, but do not compute interactions between specific input variables.

Our primary contribution is to use Sobol\'{} indices for BART model-based input activity. For a given BART MCMC draw, we derive analytic expressions that can be computed exactly for interactions of any order and do not require Monte Carlo integration, which can be expensive when the number of input variables is large. We furthermore establish a connection between first-order (main-effects) Sobol\'{} indices and one-way counts. Finally, we compare the methods using analytic test functions and demonstrate that Sobol\'{} indices applied to BART accurately capture true variable effects while remaining computationally attractive and easy to interpret. To perform this comparison, we consider both the estimation of the Sobol\'{} indices and evaluate the order-preserving sequence of active variables using a proposed novel rank-order statistic.

The rest of the paper is organized as follows. In Section \ref{sec:bartreview}, we review BART. In Section \ref{sec:sobol}, we derive Sobol\'{} indices for BART and establish a connection between first-order Sobol\'{} indices and one-way counts. In Section \ref{sec:comp}, we provide computational details. In Section \ref{sec:app}, we introduce our rank-order statistic, perform simulation studies, and apply Sobol\'{} indices to a BART-based emulator of the En-ROADS climate simulator. In Section \ref{sec:discuss}, we conclude the paper with a discussion. Proofs of stated theorems can be found in the Appendix.

%%%%%%%%%%%%%%%%%%%%%%%%%%%%%%%%%%%%%%%%%%%%%%%%%%%%%%%%%%%%%%%%%%%%%%%%%%
%%%%%%%%%%%%%%%%%%%%%%%%%%%%%%%%%%%%%%%%%%%%%%%%%%%%%%%%%%%%%%%%%%%%%%%%%%%%%%%%%%%%%%%%%%%%%%%%%%%
\section{Review of BART} 
\label{sec:bartreview}

We wish to make inference on an unknown function $f$$:$$D \rightarrow \mathbb{R}$, where domain $D$ is a $p$-dimensional subset of $\mathbb{R}^p$. We will assume for the rest of the text that domain $D$ is a bounded hyperrectangle, i.e. $D = \prod_{j=1}^p I_D^j = \prod_{j=1}^p [a_D^j, b_D^j]$, where $I_D^j$ is the $j$th marginal interval of $D$ for $j = 1, \ldots, p$. We observe the data $\dat := \{(y(\x_i), \x_i)\}_{i=1, \ldots, n}$, where each observation $y(\x)$, based on predictor $\x = (x_1, \ldots, x_p) \in D$, is assumed to be a realization of the random variable 
\begin{equation}
Y(\x) = f(\x) + \epsilon,
\label{eq:datagen}
\end{equation}
where $\epsilon \iid N(0, \sigma^2)$. 

To make inference about the unknown function $f(\bm{\cdot})$, we approximate it by a sum of $m$ regression trees. That is, we make the approximation 
\begin{equation}
f(\x) \approx \sum_{t=1}^m g(\x ; \T_t, \M_t),
\label{eq:modelSOT}
\end{equation}
where each $g(\bm{\cdot} ; \T_t, \M_t): D \rightarrow \mathbb{R}$ denotes a regression tree function and the parameters $\{ \T_t, \M_t\}_{t=1}^m$ are given a prior distribution in BART's hierarchical Bayesian model structure.
Each $g(\bm{\cdot} ; \T_t, \M_t)$ contributes a small portion to the total approximation of $f(\bm{\cdot})$. 
Hence, the expected response $\E\big[Y(\x) \mid \{(\T_t, \M_t)\}_{t=1}^m \big]$ at a given input $\x$ is the sum of each of the contributions $g(\x ; \T_t, \M_t)$. The model in Equation \ref{eq:modelSOT} is called a sum-of-trees model. 
%\blue{what is random?}

\subsection{Single-tree model}
\label{sec:modelST}

To explain the sum-of-trees model, we will first set the number of trees $m = 1$ and describe the notation of the resulting single-tree model. 

The single-tree model is the Bayesian implementation of the Classification and Regression Tree (CART) model as proposed in \cite{CGM98}. CART can be used for classification, but we assume for the paper that it is being applied to the regression setting as described in Equation \ref{eq:datagen}. The CART model provides a prediction of $f(\x)$ at input point $\x$ given observed data $\dat$. 
CART partitions the input space and fits a constant mean model in each subregion to form the predictions. 
CART constructs the partition via a binary tree structure. 
To form the partitions, each internal node contains a boolean split rule. Starting at the root node, if an input point $\x$ satisfies the split rule, it will travel to the node's left child; otherwise $\x$ will travel to the right child. The input point $\x$ will continue to traverse through the tree in this way until it reaches a terminal node.  This terminal node's parameter is the predicted value of $f(x)$.

Figure \ref{fig:tree1} shows an illustrative example. Suppose the tree in Figure \ref{fig:tree1a} is used in a single-tree model to predict an output value for input $\x^* = (x^*_1, x^*_2) = (0.9, 0.6)$, where the input space $D$ is the closed unit-square $[0, 1]^2$. Starting at the root node in Figure \ref{fig:tree1a}, we see that $\x^*$ satisfies this split rule (i.e. $x^*_2 < 0.7$), which moves $\x^*$ to the left child. We then see that $\x^*$ does not satisfy this split rule (i.e. $x^*_1 \geq 0.2$), which moves $\x^*$ to the right child, which turns out to be a terminal node. Because we are using a single-tree model (i.e. there is exactly $m=1$ tree), this mean parameter $\mu_2$ becomes the predicted value for input $\x^*$. Figure \ref{fig:tree1b} shows the corresponding hyperrectangle view of the tree.

A tree's parameters can now be organized in the following manner. 
Let $\T$ denote the set of parameters associated with the tree's split rules (i.e. the split variable and cutpoint for each internal node) and topology. 
Let $\M$ denote the set $\lbrace\mu_k\rbrace$ of parameters associated with the tree's terminal nodes.  
The single-tree model is thus $f(\bm{\cdot}) \approx g(\bm{\cdot}; \T, \M)$, where $f(\bm{\cdot})$, defined in Equation \ref{eq:datagen}, is the mean of the observed process.
Here, we think of $g(\bm{\cdot}; \T, \M)$ being a function that assigns a value $\mu_k$ to input $\x$ according to the parameters in $\T$ and $\M$.
Let $\R_k\subset D$ denote the hyperrectangle associated with the tree's terminal node that contains parameter $\mu_k$. Then,
\begin{align}
g(\bm{\cdot}; \T, \M) = \sum_{k = 1}^{|\M|} \mu_{k} \1_{\R_k} (\bm{\cdot}).
\label{eq:treebasis}
\end{align}
We may further decompose each hyperrectangle $\R_k$ into the Cartesian product of its $p$ marginal intervals $I_k^1, \ldots, I_k^p$ and hence write $\1_{\R_k} (\x) = \prod_{i=1}^p \1_{I_k^i} (x_i)$.

\subsection{Sum-of-trees model}

Now consider the sum-of-trees model in Equation \ref{eq:modelSOT} for $m > 1$. If the parameter sets $\{(\T_t, \M_t)\}_{t=1}^m$ have been established, we will let the function 
\begin{align}
\ens(\bm{\cdot}; \{\T_t, \M_t\}_{t=1}^m) := \sum_{t=1}^m g(\bm{\cdot} ; \T_t, \M_t) = \sum_{t=1}^m \sum_{k=1}^{|\M_t|} \mu_{tk} \1_{\R_{tk}} (\bm{\cdot})
\label{eq:enssot}
\end{align} 
denote the sum-of-trees approximation in Equation \ref{eq:modelSOT}. 
To streamline notation, we will refer to $\ens$ as both the function $\ens(\bm{\cdot}; \{\T_t, \M_t\}_{t=1}^m)$ and as the collection $\{(\T_t, \M_t)\}_{t=1}^m$. Thus, we write $(\T, \M) \in \ens$ if $(\T, \M) = (\T_t, \M_t)$ for some $t = 1, \ldots, m$.

\subsection{Bayesian tree models}

The sum-of-trees model is specified by the parameters $\{(\T_t, \M_t)\}_{t=1}^m$ and $\sigma^2$. Hence, a trained BART model will sample from the posterior distribution
\begin{equation}
\pi(\Theta \mid \dat) \propto L(\Theta \mid \dat) \, \pi(\Theta), 
\label{eq:BARTposterior}
\end{equation}
where $\Theta = \{(\T_1, \M_1), (\T_2, \M_2), \ldots, (\T_m, \M_m), \sigma^2\}$ are the parameters, $\dat$ is the observed data,
\begin{align*}
    L(\Theta \mid \dat) 
    &\propto \sigma^{-n} \exp\bigg(-\frac{1}{2 \sigma^2} \sum_{i=1}^n \Big(y(\x_i) - \sum_{t=1}^m g(\x_i; \T_t, \M_t)\Big)^2 \bigg) 
\end{align*}
is the likelihood, and $\pi(\Theta)$ is the prior. 

\cite{CGM10} specify the full prior $\pi(\Theta)$ by constraining it to satisfy independence conditions
\begin{equation}
\pi(\Theta) = \Bigg[ \prod_{t=1}^m \pi(\M_t \mid \T_t) \pi(\T_t) \Bigg] \pi(\sigma^2),
\label{eq:cgmprior}
\end{equation}
and
\begin{equation}
\pi(\M_t \mid \T_t) = \prod_{k=1}^{\mid\M_t\mid} \pi(\mu_{tk} \mid \T_t)
% \pi(\M_t \mid \T_t) = \prod_{\mu_{tk} \in \M_t} \pi(\mu_{tk} \mid \T_t).
\label{eq:mupriorindep}
\end{equation}
for all $t = 1, \ldots, m$. 
In Equation \ref{eq:cgmprior}, the parameter sets $(\T_t, \M_t)$ and $\sigma^2$ are constrained to be mutually independent. In Equation \ref{eq:mupriorindep}, the terminal node parameters of every tree are constrained to be independent. These independence conditions simplify the problem of specifying the full prior $\pi(\Theta)$ to specifying only the priors $\pi(\T_t)$, $\pi(\mu_{tk} \mid \T_t)$, and $\pi(\sigma^2)$. Forcing the priors $\pi(\T_t)$ and $\pi(\mu_{tk} \mid \T_t)$ to be identical for all $k=1,\dots,\vert\M_t\vert$ and $t = 1, \ldots, m$ further simplifies the prior specification problem. Furthermore, \cite{CGM98} choose the three prior forms to simplify analysis and computation by taking advantage of known conjugacy pairs. In particular, they choose the $\pi(\mu_{tk} \mid \T_t)$ prior to be a conjugate Normal distribution. To configure the priors, \cite{CGM10} recommend automatically specifying the relevant hyperparameters using data-driven methods. 

The posterior in Equation \ref{eq:BARTposterior} can thus be sampled using the following Gibbs sampler:
\begin{enumerate}
\item Draw $\{(\T_t, \M_t)\}_{t=1}^m \mid \sigma^2, \dat$. \label{item:tmSoT}
\item Draw $\sigma^2 \mid \{(\T_t, \M_t)\}_{t=1}^m, \dat$. \label{item:sigmaSoT}
\end{enumerate}
For Step \ref{item:sigmaSoT}, we can draw $\sigma^2 \mid \{(\T_t, \M_t)\}_{t=1}^m, \dat$ by performing a simple conjugate Gibbs step. Step \ref{item:tmSoT} itself will also be a Gibbs sampler that relies on being able to sample from the conditional distribution 
\begin{equation}
\pi(\T_t, \M_t \mid \{(\T_{\tau}, \M_{\tau})\}_{\tau \neq t}, \sigma^2, \dat)
\label{eq:conddistr}
\end{equation}
for all $t = 1, \ldots, m$. 
To sample from this conditional distribution, we simplify the likelihood by noting 
\begin{align*}
    L(\Theta \mid \dat) 
    &\propto \sigma^{-n} \exp\bigg(-\frac{1}{2 \sigma^2} \sum_{i=1}^n \Big(r_t(\x_i) - g(\x_i; \T_t, \M_t)\Big)^2 \bigg) 
\end{align*}
where $r_t(\mathbf{x}_i) := y(\mathbf{x}_i) - \sum_{\tau \neq t} g(\x_i; \T_{\tau}, \M_{\tau})$. 
Therefore, the conditional distribution in Equation \ref{eq:conddistr} for any $t = 1, \ldots, m$ relies on $\{(\T_{\tau}, \M_{\tau})\}_{\tau \neq t}$ and $\dat$ only through $\R_t = \{(r_t(\x_i), \x_i)\}_{i=1, \ldots, n}$. Hence, the conditional distribution can be expressed as $\pi(\T_t, \M_t \mid \R_{t}, \sigma^2)$, where $\R_t$ plays the role of $\dat$ in the single-tree version of Step \ref{item:tmSoT} of the Gibbs sampler. 
Each draw from the conditional distribution in Equation \ref{eq:conddistr} for any $t = 1, \ldots, m$ is then reduced to two draws:
\begin{enumerate}
    \item[(a)] Draw $\T_t \mid \sigma^2, \R_{t}$.
    \item[(b)] Draw $\M_t \mid \T_t, \sigma^2, \R_{t}$. 
\end{enumerate}

%%%%%%%%%%%%%%%%%%%%%%%%%%%%%%%%%%%%%%%%%%%%%%%%%%%%%%%%%%%%%%%%%%%%%%%%%%
%%%%%%%%%%%%%%%%%%%%%%%%%%%%%%%%%%%%%%%%%%%%%%%%%%%%%%%%%%%%%%%%%%%%%%%%%%%%%%%%%%%%%%%%%%%%%%%%%%%
\section{Sobol\'{} indices} 
\label{sec:sobol}

In Section \ref{sec:intro}, we introduced the idea from \cite{Sobol93} that the variance of any real-valued function defined on and square-integrable in a unit-hypercube domain can be decomposed into a sum of variance terms. 
The \cite{Sobol93} results apply when inputs $X_1, X_2, \ldots, X_p$ are continuous and mutually uncorrelated with finite interval supports.
Thus, using Equation \ref{eq:enssot}, we can decompose the variance of a BART ensemble function into a sum of terms attributed to single inputs or to interactions between sets of inputs. 

To develop our BART-based Sobol\'{} indices, we will require the following assumptions:
\begin{enumerate}[label=\textbf{A.\arabic*}]
    \item \label{cond:uncorr} $X_1, \ldots, X_p$ are mutually uncorrelated;
    \item \label{cond:pae} $X_i$'s density $\pi_i$ is positive almost everywhere on the domain's $i$th margin;
    \item \label{cond:as} Conditional on parameter sets $\{(\T_t, \M_t)\}_{t=1}^m$, $\ens(\X; \{\T_t, \M_t\}_{t=1}^m) = \ens(\x^*)$ holds if and only if input points $\x$ and $\x^*$ belong to the same set of $m$ terminal nodes.  
\end{enumerate}
We use conditions \ref{cond:uncorr} and \ref{cond:pae} to extend the two original results from \cite{Sobol93} and to derive Sobol\'{} indices for BART ensembles. 
Condition \ref{cond:as} follows from each $\mu_k| \T$ being conditionally Normal. 
If inputs $\x$ and $\x^*$ belong to different terminal nodes in at least one of the ensemble's trees, then the probability that $\ens(\X; \{\T_t, \M_t\}_{t=1}^m) = \ens(\x^*)$ is zero. 
Therefore, condition \ref{cond:as} is a reasonable assumption to make. 
We also note that condition \ref{cond:as} is used only when relating Sobol\'{} indices to counts and does not affect the computation of Sobol\'{} indices for BART ensembles.

We can now state the desired generalized version of the variance decomposition described in \cite{Sobol93}. 
For any random vector $\X = (X_1, \ldots, X_p)$ that satisfies conditions \ref{cond:uncorr} and \ref{cond:pae} on $p$-dimensional bounded hyperrectangle domain $D$ and for any real-valued function $f$ square-integrable on $D$, the variance of $f(\X)$ can be decomposed into a sum of terms attributed to single inputs or to interactions between sets of inputs. 
That is, 
\begin{align}
\Var_{\X} \big(f(\X)\big)
&= \sum_{i=1}^p V_i + \sum_{i=1}^p \sum_{i<j} V_{ij} + \bm{\cdots} + V_{1 2 \ldots p}, 
\label{eq:vardecomp}
\end{align}
where we recursively define for each variable index set $P \subseteq \{1, 2, \ldots, p\}$
\begin{align}
V_P :&= \Var_{\X_P}\Big( \E_{\X_{-P}}[f(\X) \mid \X_P] \Big) - \sum_{Q \subset P} V_Q
\label{eq:vardecompdef}
\end{align}
where the sum is over all nonempty, proper subsets $Q$ of $P$. 
In particular, the (unnormalized) first-order Sobol\'{} index $V_i$ (i.e. $V_P$ when $P = \{i\}$) is 
\begin{equation*}
V_i := \Var_{X_i}( \E_{\X_{-i}}[f(\X) \mid X_i] )
\end{equation*} 
for all $i=1, \ldots, p$. Also, the (unnormalized) second-order Sobol\'{} index $V_{ij}$ is 
\begin{equation*}
V_{ij} := \Var_{\X_{ij}}( \E_{\X_{-ij}}[f(\X) \mid \X_{ij}] ) - V_i - V_j
\end{equation*}
for all $i \neq j$. 
Often, each $V_P$ term is divided by the total variance $\Var_{\X}(f(\X))$ to produce normalized Sobol\'{} index
\begin{align}
S_P :&= V_P/\Var_{\X}(f(\X)).
\label{eq:spgen}
\end{align} 
By Equation \ref{eq:vardecomp}, the sum of all normalized indices equals unity. This allows comparison between indices of different functions.

\subsection{Sobol\'{} indices applied to BART}
\label{sec:sobolbart}

Next, we apply this variance decomposition for general $L_2$ functions $f(\bm{\cdot})$ to BART ensemble functions $\ens(\bm{\cdot}; \{\T_t, \M_t\}_{t=1}^m)$. 
That is, we will compute the terms in the right hand side of Equation \ref{eq:vardecomp} for BART ensembles. 

The core terms to compute in Equation \ref{eq:vardecompdef} are the conditional expectation $\E_{\X_{-P}}[\ens(\X; \{\T_t, \M_t\}_{t=1}^m) \mid \X_P]$ and its variance with respect to $\X_P$. 
By integrating both sides of Equation \ref{eq:enssot}, we obtain an analytic expression for the conditional expectation:
\begin{align}
\E_{\X_{-P}}[\ens(\X; \{\T_t, \M_t\}_{t=1}^m) \mid \X_P] = \sum_{k \in B_{\ens}} d_k^{-P} \1_{\R_k^P} (\X_P), 
\label{eq:ceoneway}
\end{align}
where the set $B_{\ens}$ indexes the terminal nodes of ensemble $\ens$, the $|P|$-dimensional hyperrectangle $\R_k^P$ is the projection of terminal node $k$'s $p$-dimensional hyperrectangle $\R_k$ onto the dimensions in $P$, and $d_k^{-P} = \mu_k \Prob_{-P}(\R_k^{-P})$, where we introduce the notation $\Prob_P(\bm{\cdot}) = \Prob_{\X_P}(\bm{\cdot}) = \Prob(\X_P \in \bm{\cdot})$. 
Theorem \ref{thm:sobolbart} then provides an analytic expression for the variance of the conditional expectation. 

\begin{thm}
\label{thm:sobolbart}
For any random vector $\X = (X_1, \ldots, X_p)$ that satisfies conditions \ref{cond:uncorr} and \ref{cond:pae} on a $p$-dimensional bounded hyperrectangle domain $D$, the variance of the conditional expectation in Equation \ref{eq:ceoneway} with respect to variable index set $P$ is
\begin{equation}
\Var_{\X_P}\Big( \E_{\X_{-P}}[\ens(\X; \{\T_t, \M_t\}_{t=1}^m) \mid \X_P] \Big) = \sum_{k \in B_{\ens}} \sum_{l \in B_{\ens}} d_k^{-P} d_l^{-P} C^P_{k, l}
\label{eq:Sobolkernel}
\end{equation}
where $d_k^{-P} = \mu_k \Prob_{-P}(\R_k^{-P})$ and $C^P_{k, l} = \Prob_P(\R_k^P \cap \R_l^P) - \Prob_P(\R_k^P) \Prob_P(\R_l^P)$ (if $P = \{1, \ldots, p\}$, then $d_k^{-P} = \mu_k$). In particular, the (unnormalized) first-order Sobol\'{} index $V_i$ is
\begin{equation*}
V_i = \sum_{k \in B_{\ens}} \sum_{l \in B_{\ens}} d_k^{-i} d_l^{-i} C^i_{k, l},
\end{equation*}
where $d_k^{-i} = \mu_k \prod_{j \neq i} \Prob_j(I_k^j)$ and $C^i_{k, l} = \Prob_i(I_k^i \cap I_l^i) - \Prob_i(I_k^i) \Prob_i(I_l^i)$.
\end{thm}

\subsection{How do counts and Sobol\'{} indices relate?}
\label{sec:sobolcompare}

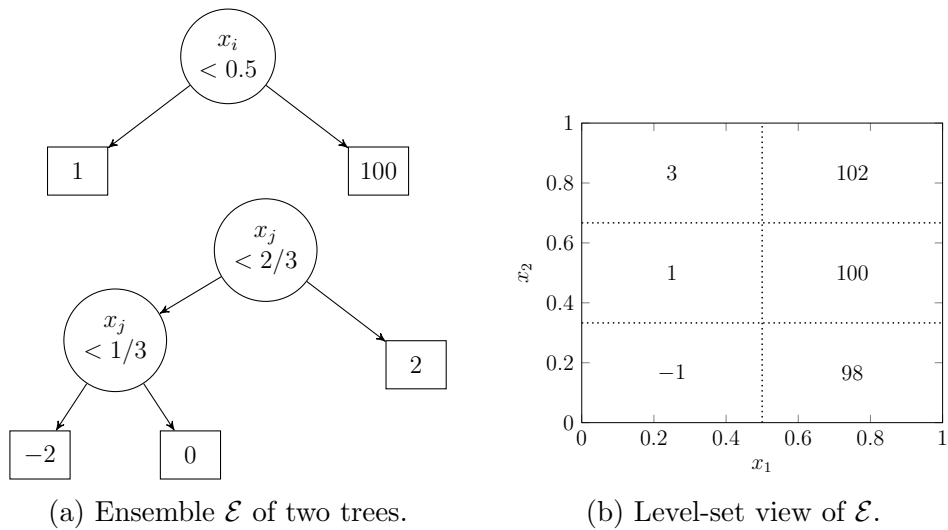
\begin{figure}[t]
    \centering
    \begin{subfigure}[b]{0.48\textwidth}
        \centering
        \begin{tikzpicture}[->,>=stealth', level/.style={sibling distance = 5cm/#1, level distance = 1.5cm}, scale=0.8,transform shape]
            \node [treenode] {$x_i$ \\ $< 0.5$}
            child { node [terminal] {$1$} }
            child { node [terminal] {$100$} }
        ;
        \end{tikzpicture}

        \begin{tikzpicture}[->,>=stealth', level/.style={sibling distance = 5cm/#1, level distance = 1.5cm}, scale=0.8,transform shape]
            \node [treenode] {$x_j$ \\ $< 2/3$}
            child { 
                node [treenode] {$x_j$ \\ $< 1/3$} 
                child { node [terminal] {$-2$} }
                child { node [terminal] {$0$} }
            }
            child { node [terminal] {$2$} }
        ;
        \end{tikzpicture}
        \caption{Ensemble $\ens$ of two trees.}
        \label{fig:ordera}
    \end{subfigure}
    \begin{subfigure}[b]{0.48\textwidth}
        \centering
        \begin{tikzpicture}[scale=0.7]
            \begin{axis}[xmin=0, xmax=1, ymin=0, ymax=1, xlabel={$x_1$}, ylabel={$x_2$}]
                \addplot [dotted, thick] table {
                    0.5 0
                    0.5 1
                };
                \addplot [dotted, thick] table {
                    0 0.333
                    1 0.333
                };
                \addplot [dotted, thick] table {
                    0 0.667
                    1 0.667
                };
                \node at (axis cs:0.25,0.167) {$-1$};
                \node at (axis cs:0.25,0.5) {$1$};
                \node at (axis cs:0.25,0.833) {$3$};
                \node at (axis cs:0.75,0.167) {$98$};
                \node at (axis cs:0.75,0.5) {$100$};
                \node at (axis cs:0.75,0.833) {$102$};
            \end{axis}
        \end{tikzpicture}
        \caption{Level-set view of $\ens$.}
        \label{fig:orderb}
    \end{subfigure}
    \caption{Two different views of the same ensemble $\ens$.}
    \label{fig:order}
\end{figure}

To this day, count-based methods remain the most popular ways of assessing input activity for BART. But as \cite{LRW18} note, they are not theoretically well-understood. We have seen in Figure \ref{fig:count} a scenario in which the one-way count metric not only inaccurately measures input activity in the data-generating function $f$ but also incorrectly ranks the variables in order of importance. \cite{CGM10} and \cite{Bleich14} also detail scenarios that question how accurately one-way count metric assess input activity in the data-generating function and suggest ad-hoc work-arounds, such as fitting BART with small $m$ to get an empirically better behaved estimate of input activity.
But how do counts perform when assessing input activity in the BART ensemble itself? To answer this question, we turn to the example in Figure \ref{fig:ordera}. The count metric will look at number of splits and conclude that variable $x_j$ is twice as active than variable $x_i$. But if we look at the terminal node values of the ensemble, variable $x_i$ is clearly more important than variable $x_j$ in determining the ensemble's predicted value. If the count metric is not measuring variable importance in the ensemble, then what exactly does it measure? Theorem \ref{thm:mec} answers this question.

\begin{thm}
\label{thm:mec}
Let $\ens$ be a BART ensemble of $m$ regression trees with parameters $\{\T_t, \M_t\}_{t=1}^m$. Assume $\ens$ satisfies assumptions \ref{cond:uncorr}, \ref{cond:pae}, and \ref{cond:as}, and fix $i \in \{1, \ldots, p\}$. Then the number of unique split rules in $\ens$ that involve variable $x_i$ equals the number of jumps in the piecewise-constant function $\E_{\X_{-i}}[\ens(\X; \{\T_t, \M_t\}_{t=1}^m) \mid X_i = \bm{\cdot}]$.
\end{thm}

To see why Theorem \ref{thm:mec} might be true, consider a BART ensemble $\ens_0$ with $m$ regression trees, where each tree is simply a terminal node. 
The ensemble $\ens_0$, which predicts the same value for any input $\x \in D$, can turn into any $m-$tree BART ensemble $\ens$ by undertaking an appropriate sequence of birth processes. 
Any birth process slices a terminal node's corresponding hyperrectangle into two smaller hyperrectangles according to some split rule. If we call this split rule ``$x_i < c$'', then this slice occurs on the $(p-1)$-dimensional hyperplane $x_i = c$ in $D$. The resulting ``left'' (``right'') hyperrectangle gains a terminal node parameter $\mu_{left}$ (parameter $\mu_{right}$), where $\mu_{left} \neq \mu_{right}$ through assumption \ref{cond:as}.
If the split rule ``$x_i < c$'' did not already exist in the ensemble, then the piecewise-constant function $\E_{\X_{-i}}[\ens(\X; \{\T_t, \M_t\}_{t=1}^m) \mid X_i = \bm{\cdot}]$ must have been constant at $x_i = c$ prior to the birth process, which means the birth process produces a jump in the piecewise-constant function at $x_i = c$. 
Meanwhile, no jumps are produced in any of the other piecewise-constant functions $\E_{\X_{-j}}[\ens(\X; \{\T_t, \M_t\}_{t=1}^m) \mid X_j = \bm{\cdot}]$ (where $j \neq i$). 
Hence, under assumptions \ref{cond:uncorr}, \ref{cond:pae}, and \ref{cond:as}, each birth process that produces a unique split rule that involves variable $x_i$ increments the number of jumps in the piecewise-constant function $\E_{\X_{-i}}[\ens(\X; \{\T_t, \M_t\}_{t=1}^m) \mid X_i = \bm{\cdot}]$ by one. 

Theorem \ref{thm:mec} also provides a link between the one-way count metric and the theoretically more well-understood first-order Sobol\'{} index. Under the conditions of Theorem \ref{thm:mec}, the one-way count of variable $x_i$ is the number of jumps in the conditional expectation function $\E_{\X_{-i}}[\ens(\X; \{\T_t, \M_t\}_{t=1}^m) \mid X_i = \bm{\cdot}]$.
Under the conditions of Theorem \ref{thm:sobolbart}, the first-order Sobol\'{} index of variable $x_i$ is the variance of the conditional expectation $\E_{\X_{-i}}[\ens(\X; \{\T_t, \M_t\}_{t=1}^m) \mid X_i]$.
Thus, under certain conditions, both the one-way count and the first-order Sobol\'{} index of variable $x_i$ are functions of the conditional expectation function $\E_{\X_{-i}}[\ens(\X; \{\T_t, \M_t\}_{t=1}^m) \mid X_i = \bm{\cdot}]$. 

Interestingly, the number of jumps and variance can each be viewed as a measure of variability. Under this lens, the one-way count metric can been seen as a more crude version of the first-order Sobol\'{} index. Theorem \ref{thm:mew} describes how to ``standardize'' the conditional expectation function so that its variance becomes the number of jumps of the conditional expectation. We use the term standardize because many different conditional expectation functions can be transformed into the standardized conditional expectation function, but the standardized conditional expectation function cannot be transformed back into the original conditional expectation function. 

\begin{thm}
\label{thm:mew}
Let $\ens$ be a BART ensemble satisfying \ref{cond:uncorr}, \ref{cond:pae}, and \ref{cond:as}. Recall that for all dimensions $i = 1, \ldots, p$, the conditional expectation function $\E_{\X_{-i}}[\ens(\X; \{\T_t, \M_t\}_{t=1}^m) \mid X_i = \bm{\cdot}]$ is piecewise constant and hence can be written as $\E_{\X_{-i}}[\ens(\X; \{\T_t, \M_t\}_{t=1}^m) \mid X_i = \bm{\cdot}] = \sum_{k^* \in B_{\ens}^i} e_{k^*}^i \1_{\I_{k^*}^i}(\bm{\cdot})$, where $B_{\ens}^i$ indexes the intervals of this piecewise constant function. Suppose for any indices $k^*, l^* \in B_{\ens}^i$ that $e_{k^*}^i = e_{l^*}^i$ implies $\I_{k^*}^i = \I_{l^*}^i$ (i.e. the piecewise constant function has distinct values in different input regions). Consider the following transformations to this conditional expectation function: 
\begin{enumerate}
    \item Center and scale $\{e_{k^*}^i : k^* \in B_{\ens}^i\}$ so that the corrected sample variance\footnote{The definition we use for the corrected sample variance of real numbers $x_1, \ldots, x_n$ is $(n-1)^{-1} \sum_{i=1}^n (x_i - \bar{x})^2$, where $\bar{x} = n^{-1} \sum_{i=1}^n x_i$ is the sample mean of $x_1, \ldots, x_n$.} equals $|B_{\ens}^i|$. 
    \item Assign equal probability mass $|B_{\ens}^i|^{-1}$ to each $\I_{k^*}^i$. 
\end{enumerate}
Then the number of jumps in this transformed conditional expectation function equals its variance.
\end{thm}

As with condition \ref{cond:as}, the added ``$e_{k^*}^i = e_{l^*}^i$ implies $\I_{k^*}^i = \I_{l^*}^i$'' assumption in Theorem \ref{thm:mew} follows from each $\mu_k| \T$ being conditionally Normal. We may use reasoning similar to before to argue that this assumption is also reasonable to make.

%%%%%%%%%%%%%%%%%%%%%%%%%%%%%%%%%%%%%%%%%%%%%%%%%%%%%%%%%%%%%%%%%%%
%%%%%%%%%%%%%%%%%%%%%%%%%%%%%%%%%%%%%%%%%%%%%%%%%%%%%%%%%%%%%%%%%%%%%%%%%%%%%%%%%%%%%%%%%%%%%%%%%%%
\section{Computational Details} 
\label{sec:comp}

Given a $L^2$ function $f$, we wish to estimate its normalized Sobol\'{} indices $S_P^f$ as defined in Equation \ref{eq:spgen} for all variable index sets $P$. We do so by first training a BART model on data generated from Equation \ref{eq:datagen} and drawing $N$ samples $\Theta^{(1)}, \ldots, \Theta^{(N)}$ from the resulting BART posterior in Equation \ref{eq:BARTposterior}. For each variable index set $P$, we then compute 
\begin{align*}
S_P^{\Theta^{(j)}}
:&= \frac{\Var_{\X_P}( \E_{\X_{-P}}[\ens(\X; \{\T_t^{(j)}, \M_t^{(j)}\}_{t=1}^m) \mid \X_P] ) - \sum_{Q \in 2^P \setminus \{\emptyset, P\}} V_Q^{\Theta^{(j)}}}{\Var_{\X}( \ens(\X; \{\T_t^{(j)}, \M_t^{(j)}\}_{t=1}^m) )}
\end{align*}
for each posterior draw $\Theta^{(j)}$, where $j = 1, \ldots, N$. We can then obtain a point estimate of $S_P^f$ by approximating the integral $\int S_P^{\Theta} \, d\pi(\Theta \mid \dat)$ using the sample mean of $S_P^{\Theta^{(1)}}, \ldots, S_P^{\Theta^{(N)}}$. That is, our point estimate of $S_P^f$ is 
\begin{equation*}
\hat{S}_P^f = \frac{1}{N} \sum_{j=1}^N S_P^{\Theta^{(j)}}.
\end{equation*}

At the core of these calculations is the variance term 
\[
    \Var_{\X_P}\Big( \E_{\X_{-P}}[\ens(\X; \{\T_t, \M_t\}_{t=1}^m) \mid \X_P] \Big)
\]
which we showed can be computed exactly using Theorem \ref{thm:sobolbart}. 
Furthermore, it turns out that possibly many, if not all, of the summands in Equation \ref{eq:Sobolkernel} are zero. 
Theorem \ref{thm:rmterms} below explains under what conditions a summand vanishes.

\subsection{Unnormalized Sobol\'{} indices}
% \subsection{Irrelevant terminal nodes}

A sensible goal in sensitivity analysis is to compute all first-order Sobol\'{} indices. According to Equation \ref{eq:Sobolkernel}, each unnormalized first-order index $V_i$ for $\ens$ is a sum of $|B_{\ens}|^2$ terms. Hence, computing all $p$ first-order indices requires calcuating $p\times|B_{\ens}|^2$ terms. However, we may take advantage of BART's additive structure to reduce the number of terms to compute. 

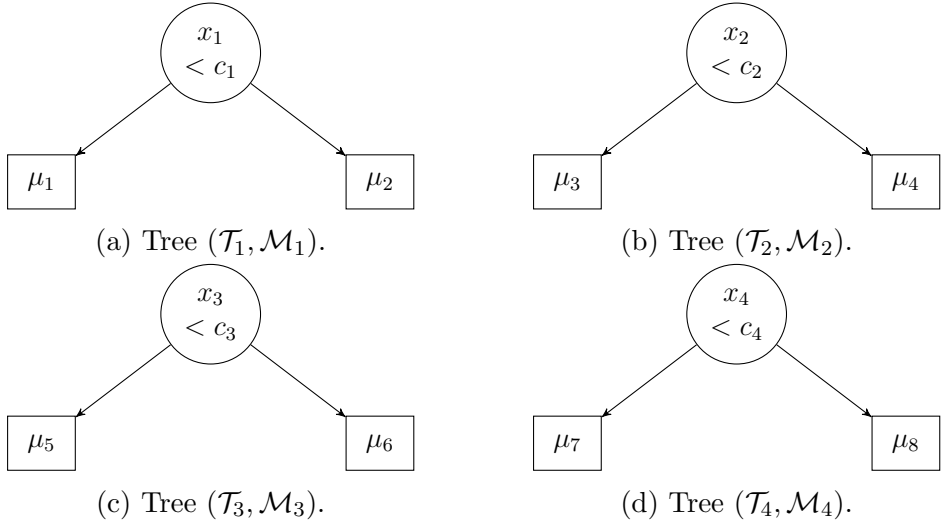
\begin{figure}[t]
\centering
\begin{subfigure}[b]{0.48\textwidth}
\centering
    \begin{tikzpicture}[->,>=stealth', level/.style={sibling distance = 5cm/#1, level distance = 1.5cm}, scale=0.9, transform shape]
    \node [treenode] {$x_1$ \\ $< c_1$}
    child { node [terminal] {$\mu_1$} }
    child { node [terminal] {$\mu_2$} }
;
\end{tikzpicture}
\caption{Tree $(\T_1, \M_1)$.}
\end{subfigure}
~
\begin{subfigure}[b]{0.48\textwidth}
\centering
    \begin{tikzpicture}[->,>=stealth', level/.style={sibling distance = 5cm/#1, level distance = 1.5cm}, scale=0.9, transform shape]
    \node [treenode] {$x_2$ \\ $< c_2$}
    child { node [terminal] {$\mu_3$} }
    child { node [terminal] {$\mu_4$} }
;
\end{tikzpicture}
\caption{Tree $(\T_2, \M_2)$.}
\end{subfigure}
~
\begin{subfigure}[b]{0.48\textwidth}
\centering
    \begin{tikzpicture}[->,>=stealth', level/.style={sibling distance = 5cm/#1, level distance = 1.5cm}, scale=0.9, transform shape]
    \node [treenode] {$x_3$ \\ $< c_3$}
    child { node [terminal] {$\mu_5$} }
    child { node [terminal] {$\mu_6$} }
;
\end{tikzpicture}
\caption{Tree $(\T_3, \M_3)$.}
\end{subfigure}
~
\begin{subfigure}[b]{0.48\textwidth}
\centering
    \begin{tikzpicture}[->,>=stealth', level/.style={sibling distance = 5cm/#1, level distance = 1.5cm}, scale=0.9, transform shape]
    \node [treenode] {$x_4$ \\ $< c_4$}
    child { node [terminal] {$\mu_7$} }
    child { node [terminal] {$\mu_8$} }
;
\end{tikzpicture}
\caption{Tree $(\T_4, \M_4)$.}
\end{subfigure}
\caption{Four trees.}
\label{fig:meex1}
\end{figure}

Consider the example ensemble $\ens$ consisting only of the $m=4$ trees in Figure \ref{fig:meex1}. Because ensemble $\ens$ has $|B_{\ens}| = 8$ terminal nodes, the unnormalized first-order Sobol\'{} index $V_1$ is a sum of $|B_{\ens}|^2 = 64$ terms.
However, only tree $(\T_1, \M_1)$ splits on variable $x_1$, which makes the conditional expectations $\E[g(\X; \T_2, \M_2) \mid X_1 = x_1]$, $\E[g(\X; \T_3, \M_3) \mid X_1 = x_1]$, and $\E[g(\X; \T_4, \M_4) \mid X_1 = x_1]$ constant in $x_1$. Thus, 
\begin{align*}
V_1 &= \Var_{X_1}\big( \E[\ens(\X; \{\T_t, \M_t\}_{t=1}^m) \mid X_1] \big) \\
&= \Var_{X_1}\Bigg( \E\Big[\sum_{t=1}^m g(\X; \T_t, \M_t) \mid X_1\Big] \Bigg) \\
&= \Var_{X_1}\big( \E[g(\X; \T_1, \M_1) \mid X_1] \big),
\end{align*}
which, according to Equation \ref{eq:Sobolkernel}, is a sum of only $|\M_1|^2 = 4$ terms. Using the same logic, each Sobol\'{} index $V_2$, $V_3$, and $V_4$ reduces from a sum of $|B_{\ens}|^2 = 64$ terms to a sum of, respectively, $|\M_2|^2 = 4$, $|\M_3|^2 = 4$, and $|\M_4|^2 = 4$ terms. Hence, computing all four indices $V_1$, $V_2$, $V_3$, and $V_4$ reduces from a sum of $4\times|B_{\ens}|^2 = 256$ terms to a sum of $|\M_1|^2 + |\M_2|^2 + |\M_3|^2 + |\M_4|^2 = 16$ terms.

More generally, to compute Equation \ref{eq:Sobolkernel} for arbitrary variable index set $P$, we may remove any tree that does not split on any variable in $P$. Furthermore, we may take advantage of the ensemble function's formulation in Equation \ref{eq:enssot} to remove any node whose path to root node does not split on any variable in $P$. This statement is made precise in Theorem \ref{thm:rmterms}. 

\begin{thm}
\label{thm:rmterms}
Let $\ens(\bm{\cdot}; \{\T_t, \M_t\}_{t=1}^m) = \sum_{k \in B_{\ens}} \mu_{k} \1_{\R_k} (\bm{\cdot})$ be the function of a BART ensemble $\ens$. 
For any terminal node $\eta_k$, let $v(k)$ be the index set of all split variables along the path to $\eta_k$'s root node. 
For any variable index set $P$, let 
\begin{equation*}
\ens_P(\bm{\cdot}; \{\T_t, \M_t\}_{t=1}^m) := \sum_{\substack{k \in B_{\ens} \\ v(k) \cap P \neq \emptyset}} \mu_{k} \1_{\R_k} (\bm{\cdot})
\end{equation*} 
be the ensemble function that results from removing from $B_{\ens}$ any terminal node whose path to root node does not split on any variable in $P$. Then 
\begin{equation*}
\Var_{\X_P}( \E[\ens(\X; \{\T_t, \M_t\}_{t=1}^m) \mid \X_P] ) = \Var_{\X_P}( \E[\ens_P(\X; \{\T_t, \M_t\}_{t=1}^m) \mid \X_P] ).
\end{equation*}
\end{thm}

To get a better sense of how much computation Theorem \ref{thm:rmterms} saves, consider again the goal of computing all $p$ (unnormalized) first-order Sobol\'{} indices if an ensemble $\ens$, but for a more realistic scenario: 
let $p = 10$ and suppose the ensemble $\ens$ is such that $N_1 = N_2 = N_3 = N_4 = N_5 = \frac{1}{4} |B_{\ens}|$ and $N_6 = N_7 = N_8 = N_9 = N_{10} = 0$, where $N_i$ is the number of terminal nodes in $\ens$ whose path to root node includes split variable $x_i$. 
Theorem \ref{thm:rmterms} tells us for all $i = 1, \ldots, p$, the first-order index $V_i$ is a sum of $N_i^2$ terms. Hence, computing all $p$ first-order Sobol\'{} indices would require $\sum_{i=1}^p N_i^2 = 5 \times \frac{1}{16} |B_{\ens}|^2$ terms to be computed, which is a $32$-fold improvement over $p\times|B_{\ens}|^2$, which is the number if terms to compute if Theorem \ref{thm:rmterms} is not used.

\subsection{Total-effects index}

In sensitivity analysis, we also often wish to obtain some measure of interaction between the input variables. We can do so via the total-effects sensitivity index, which is defined to be the sum of all normalized sensitivity indices involving the input variable in question \citep{Saltelli00}. For example, if $p = 3$, then the total-effects index for input variable $x_2$ would be $T_2 = S_2 + S_{12} + S_{23} + S_{123}$. Hence, $T_2 - S_2$ provides a sense of the magnitude of all interactions involving variable $x_2$. However, in order to compute all $p$ total-effects sensitivity indices, this formulation requires computing all $2^p - 1$ normalized sensitivity indices $S_P$. Fortunately, the total-effects index is equivalent to the following expression: 
\begin{equation*}
T_i = 1 - S_{-i},
\end{equation*}
where $S_{-i} = S_{\{1, \ldots, i-1, i+1, \ldots, p\}}$, and $\sum_{i=1}^p T_i \geq 1$ with equality only if the model is purely additive.  
Hence, we only need to compute $p$ of these variance expressions in order to compute all $p$ total-effects sensitivity indices.

%%%%%%%%%%%%%%%%%%%%%%%%%%%%%%%%%%%%%%%%%%%%%%%%%%%%%%%%%%%%%%%%%%%%%%%%%%%%
%%%%%%%%%%%%%%%%%%%%%%%%%%%%%%%%%%%%%%%%%%%%%%%%%%%%%%%%%%%%%%%%%%%%%%%%%%%%%%%%%%%%%%%%%%%%%%%%%%%
\section{Applications} 
\label{sec:app}

\subsection{Simulation study}

\subsubsection*{Simulation settings discussion.} 
Given data generated from Equation \ref{eq:datagen}, where the true Sobol\'{} index values for $f(\x)$ are known, this section identifies the number of inputs, the sample size, and the magnitude of the measurement error standard deviation which answer the following questions: 
\begin{enumerate}[label=\textbf{Q.\arabic*}]
    \item 
    \label{qu:accuracy}  
    What is the bias of BART-based Sobol\'{} indices for estimating the Sobol\'{} indices of $f(\bm{\cdot})$ when $f(\x)$ is measured with error?
    \item 
    \label{qu:accuracy2}  
    How does this bias compare to the bias of estimated Sobol\'{} indices obtained using alternative nonparametric prediction models?
    \item 
    \label{qu:ranking}  
    How close are the first-order rankings provided by BART-based Sobol\'{}  indices to the first-order ranking provided by the Sobol\'{} indices of $f(\bm{\cdot})$?
    \item 
    \label{qu:countrankings} 
    How close are the first-order rankings provided by one-way BART counts to the first-order ranking provided by the Sobol\'{} indices of $f(\bm{\cdot})$?
\end{enumerate}

For questions \ref{qu:accuracy}, \ref{qu:ranking}, and \ref{qu:countrankings}, and for each simulation setting below, we generate a $n-$point maximin LHD sample on $[0, 1]^p$ \citep{lhs}. 
At each input point $\x$ of the LHD sample, we generate response values for a given data-generating function $f(\bm{\cdot})$$:$$[0, 1]^p \rightarrow \mathbb{R}$ from Equation \ref{eq:datagen} where $\epsilon \sim N(0, \sigma^2)$.
We generate 500 data sets for each possible combination of the following different parameter settings: $p/p_0 \in \{1, 2, 3\}$, $n \in \{10p, 50p\}$, and $\sigma^2 \in \{0.1 \Var(f(\X)), 0.25 \Var(f(\X))\}$, where $(p-p_0)$ is the number of inert variables in $f(\bm{\cdot})$ and $\X = (X_1, \ldots, X_p)$, where each $X_i \iid U(0, 1)$. For each $f(\bm{\cdot})$, the variance $\Var(f(\X))$ is calculated analytically where possible, otherwise numerical integral approximations are used. To each of these 500 data sets, we fit a BART model using the default parameter settings of the Open Bayesian Trees (OpenBT) project found at \url{https://bitbucket.org/mpratola/openbt/} \citep{openbt}. 

The five data-generation functions to be examined are:

\begin{enumerate}
    \item From \cite{Friedman91}, the data-generating function is defined as \[f(\x) = 10\sin(\pi x_1 x_2) + 20(x_3 - 0.5)^2 + 10x_4 + 5x_5.\] This function, used in \cite{CGM10} and many other BART papers for variable activity and selection, is a challenging mix of interactions and nonlinearities. Here, only $p_0 = 5$ variables influence the response. This function's first-order Sobol\'{} indices and total-effects Sobol\'{} indices have the same ordering. That is, $S_4^f > S_1^f = S_2^f > S_3^f > S_5^f$ and $T_4^f > T_1^f = T_2^f > T_3^f > T_5^f$. Also, $\Var(f(\X)) \approx 23.8$.
    \item We modify the Friedman function to create the data-generating function defined as \[f(\x) = 10\sin(\pi (x_1 - 0.5) (x_2-0.5)) + 20(x_3 - 0.5)^2 + 10x_4 + 5x_5.\] 
    For this modified version, the first-order Sobol\'{} indices for variables 1 and 2 are zero, which changes the order of the first-order Sobol\'{} indices while maintaining the total-effects order (see Table \ref{tbl:VA1}). Also, $\Var(f(\X)) \approx 19.0$.
    \item The $g-$function from \cite{Saltelli95} with $p_0$ inputs is defined to be \[f(x_1, \ldots, x_{p_0}) = \prod_{k=1}^{p_0} \frac{|4x_k-2| + c_k}{1+c_k},\] where $\mathbf{c} = (c_1, \ldots, c_{p_0})$ has nonnegative components. This function is a product of univariate functions, which presents a greater challenge to BART than sums of univariate or bivariate functions provide. Here we use the coefficient values $c_k = (k-1)/2$ for $k = 1, \ldots, p_0$ suggested by \cite{Crestaux09}. We also use $p_0 = 5$ active variables which gives us $\Var(f(\X)) \approx 3.076$. 
    \item The Bratley function \citep{Bratley1992,Kucherenko2011} with $p_0$ inputs is defined to be 
    \begin{align*}
        f(\x) :&= \sum_{i=1}^{p_0} (-1)^i \prod_{j=1}^i x_j \\
        &= -x_1 + x_1 x_2 - x_1 x_2 x_3 + x_1 x_2 x_3 x_4 - x_1 x_2 x_3 x_4 x_5 + \cdots
    \end{align*}
    This function is a sum of products of inputs, which provides an even greater challenge to BART than the $g-$function provides.
    Furthermore, this function produces nonzero Sobol' indices for all (non-null) variable index sets.
    Again, we use $p_0 = 5$ active variables which gives us $\Var(f(\X)) \approx 0.057$. 
    \item The function inspired by \cite{Morris2006} is defined for $p_0=5$ as 
        \begin{align*}
        f(\x) :&= \alpha \sum_{i=1}^{p_0} x_i + \beta \sum_{i=1}^{p_0-1} x_i \sum_{j=i+1}^{p_0}  x_j \\
        &= \alpha x_1 + \beta (x_1 x_2 + x_1 x_3 + x_1 x_4 + x_1 x_5) \\
        &\hspace{1mm}+ \alpha x_2 + \beta (x_2 x_3 + x_2 x_4 + x_2 x_5) \\ 
        &\hspace{1mm}+ \alpha x_3 + \beta (x_3 x_4 + x_3 x_5) \\
        &\hspace{1mm}+ \alpha x_4 + \beta (x_4 x_5) \\
        &\hspace{1mm}+ \alpha x_5
        \end{align*}
        where $\alpha = \sqrt{12} - 6\sqrt{0.1 (p_0-1)} \approx -0.331$ and $\beta = \frac{12}{\sqrt{10(p_0-1)}} \approx 1.897$ are chosen so that $S_1 = S_2 = \cdots = S_5 \approx 0.05$ and total-effects indices $T_1 = T_2 = \cdots = T_5 \approx 0.35$. Here, $\Var(f(\X)) \approx 5.25$. See also \url{www.sfu.ca/~ssurjano/morretal06.html}.
\end{enumerate}

Hence, we consider $(3 \times 2 \times 2) \times 5 = 60$ possible combinations of $(p, n, \sigma^2)$ parameter settings and data-generating functions. We will call these the 60 simulation scenarios. 

For question \ref{qu:accuracy2}, Treed Gaussian Processes (TGP) provide a benchmark to compare exact BART-based Sobol\'{} indices against \citep{Gramacy2005,Gramacy2008}. However, there exists no literature on the exact computation of TGP-based Sobol\'{} indices. Hence, we estimate first-order and total-effects TGP-based Sobol\'{} indices using the \texttt{tgp::sens()} R function \citep{Gramacy2007,Gramacy2010}. This function uses Monte Carlo integration whose error depends on its Latin hypercube design (LHD) sampling scheme. The $p+2$ required LHD samples are randomly drawn for every MCMC iteration of the TGP model fitting, which propagates the Monte Carlo integration error into the posterior variability of the indices. To minimize this error, we use $100p-$point LHD samples.

\begin{table}
\tiny
\centering
\begin{tabular}{|r || *9{c|} *6{c|}} 
\hline 
% & \multicolumn{3}{c|}{Friedman} & \multicolumn{3}{c|}{Modified Friedman} & \multicolumn{3}{c|}{$g-$function} & \multicolumn{3}{c|}{Bratley} & \multicolumn{3}{c|}{Morris} \\ \cline{2-16} 
% $i$ & $S_i^f$ & $T_i^f$ & $T_i^f - S_i^f$ & $S_i^f$ & $T_i^f$ & $T_i^f - S_i^f$ & $S_i^f$ & $T_i^f$ & $T_i^f - S_i^f$ & $S_i^f$ & $T_i^f$ & $T_i^f - S_i^f$ & $S_i^f$ & $T_i^f$ & $T_i^f - S_i^f$ \\
& \multicolumn{3}{c|}{Friedman} & \multicolumn{3}{c|}{Modified Friedman} & \multicolumn{3}{c|}{$g-$function} & \multicolumn{3}{c|}{Bratley} & \multicolumn{3}{c|}{Morris} \\ \cline{2-16} 
& $S_i^f$ & $T_i^f$ & $T_i^f -$ & $S_i^f$ & $T_i^f$ & $T_i^f -$ & $S_i^f$ & $T_i^f$ & $T_i^f -$ & $S_i^f$ & $T_i^f$ & $T_i^f -$ & $S_i^f$ & $T_i^f$ & $T_i^f -$\\ 
$i$ &  &  & $S_i^f$ &  &  & $S_i^f$ &  &  & $S_i^f$ &  &  & $S_i^f$ &  &  & $S_i^f$ \\
\hline \hline
1  & 0.197 & 0.274 & 0.077 & 0     & 0.335 & 0.335 & 0.433 & 0.701 & 0.268 & 0.688 & 0.766 & 0.078 & 0.190 & 0.210 & 0.019 \\
2  & 0.197 & 0.274 & 0.077 & 0     & 0.335 & 0.335 & 0.108 & 0.284 & 0.176 & 0.142 & 0.220 & 0.078 & 0.190 & 0.210 & 0.019 \\
3  & 0.093 & 0.093 & 0     & 0.117 & 0.117 & 0     & 0.048 & 0.135 & 0.087 & 0.051 & 0.099 & 0.048 & 0.190 & 0.210 & 0.019 \\
4  & 0.350 & 0.350 & 0     & 0.438 & 0.438 & 0     & 0.027 & 0.078 & 0.051 & 0.006 & 0.018 & 0.012 & 0.190 & 0.210 & 0.019 \\
5  & 0.087 & 0.087 & 0     & 0.110 & 0.110 & 0     & 0.017 & 0.050 & 0.033 & 0.006 & 0.018 & 0.012 & 0.190 & 0.210 & 0.019 \\
\hline
\end{tabular}
\caption{$S_i^f$, $T_i^f$, and $T_i^f - S_i^f$ for various data-generating functions $f$ and variable indices $i$.}
\label{tbl:VA1}
\end{table}

\subsection{Performance metrics}

We evaluate our results in terms of two metrics: the $L_1$ performance metric and a rank-based performance metric.

\subsubsection*{$L_1$ performance metric}
To answer question \ref{qu:accuracy} posed at the beginning of the section, we will first estimate the expectation of the $L_1$ distance $d_{L_1}(\bm{\cdot}, \bm{\cdot})$ between BART-based Sobol\'{} indices and the true Sobol\'{} indices with respect to the BART posterior $\pi(\Theta \mid \dat)$ from Equation \ref{eq:BARTposterior}. For example, if we are assessing the bias of BART-based first-order Sobol\'{} indices for a given number of inputs, sample size, and magnitude of the measurement error standard deviation (i.e for a given $(p, n, \sigma^2)$), we will estimate the expectation 
\begin{equation}
    \int d_{L_1}(\bm{S}^{\ens}, \bm{S}^f) \, d\pi(\Theta \mid \dat) 
    \approx \frac{1}{1000} \sum_{i=1}^{1000} d_{L_1}(\bm{S}^{\ens^{(i)}}, \bm{S}^f)
    \label{eq:expectation}
\end{equation}
using 1,000 posterior samples $\{(\Theta^{(i)} \mid \dat)\}_{i=1}^{1000}$, where the vectors $\bm{S}^{\ens} = (S_1^{\ens}, S_2^{\ens}, \ldots, S_p^{\ens})$ and $\bm{S}^f = (S_1^f, S_2^f, \ldots, S_p^f)$ contain all first-order Sobol\'{} indices of, respectively, BART ensemble function $\ens(\bm{\cdot}; \{\T_t, \M_t\}_{t=1}^m)$ and data-generating function $f(\bm{\cdot})$. Here, $\ens(\bm{\cdot}; \{\T_t, \M_t\}_{t=1}^m)$ is the BART ensemble function that results from posterior sample $(\Theta \mid \dat)$ while each $\ens^{(i)}(\bm{\cdot})$ is similarly the BART ensemble function that results from the $i$th posterior sample $(\Theta^{(i)} \mid \dat)$. We will make similar estimates for two-way and total-effects Sobol\'{} index calculations. 
Finally, we average over replicated data sets $\dat$ to arrive at our overall estimated $L_1$ distance.
In the example above, we will generate 500 values of the expected $L_1$ distance estimate. The sample mean and standard deviation of these 500 estimates are shown in Table \ref{tbl:raw} (the numerical results are discussed in Section \ref{sec:simres}). 

To answer question \ref{qu:accuracy2}, we follow a methodology similar to that for answering question \ref{qu:accuracy}. We highlight three key differences. First, we compute TGP-based Sobol\'{} indices, which are approximated for any given posterior sample of the trained TGP. Second, we estimate TGP-based Sobol\'{} indices for only 40 data sets due to the substantial added computational demands of the required integral approximations. Table 2 reports the sample mean and standard of these 40 estimates.
Third, we do not compute TGP-based second-order Sobol\'{} indices (i.e. $\hat{S}_{ij}^{TGP}$) because the \texttt{tgp::sens()} R function does not easily lend itself to such calculations. Hence, we rely on TGP-based total-indices to capture input variable interactions. Regarding the latter two key differences, we emphasize that TGP serves merely as a benchmark and not as the focus of this paper. Hence, we use TGP as only one possible popular alternative to BART-based Sobol\'{} indices.

\begin{table}
\centering
% \scriptsize
\tiny
\begin{tabular}{|c | r || *5{l|}}
\cline{2-7}
\multicolumn{1}{c|}{\multirow{1}{*}{}} & & \multicolumn{3}{c|}{BART: mean (sd) of 500 replicates} & \multicolumn{2}{c|}{TGP: mean (sd) of 40 replicates} \\ \cline{3-7}
\multicolumn{1}{c|}{\multirow{1}{*}{}} &  ($p$, $n$, $\sigma^2$) & $S_i^f$ vs $S_i^{\ens}$ & $S_{ij}^f$ vs $S_{ij}^{\ens}$ & $T_i^f$ vs $T_i^{\ens}$ & $S_i^f$ vs $\hat{S}_i^{TGP}$ & $T_i^f$ vs $\hat{T}_i^{TGP}$  \\ \cline{2-7} \cline{2-7} \hline\hline
\multirow{12}{*}{\rotatebox[origin=c]{90}{Friedman function}} 
&  (5, 50$p$, 0.10) & 0.072 (0.021) & 0.067 (0.006) & 0.137 (0.037) & 0.198 (0.007) & 0.444 (0.030) \\
&  (5, 50$p$, 0.25) & 0.099 (0.034) & 0.082 (0.005) & 0.174 (0.047) & 0.269 (0.019) & 0.938 (0.108) \\
&  (5, 10$p$, 0.10) & 0.184 (0.065) & 0.089 (0.001) & 0.267 (0.095) & 0.289 (0.048) & 0.886 (0.218) \\
&  (5, 10$p$, 0.25) & 0.206 (0.077) & 0.091 (0.002) & 0.289 (0.095) & 0.339 (0.058) & 1.272 (0.275) \\ \cline{2-7}
& (10, 50$p$, 0.10) & 0.074 (0.018) & 0.050 (0.006) & 0.137 (0.024) & 1.030 (0.009) & 1.100 (0.027) \\
& (10, 50$p$, 0.25) & 0.120 (0.026) & 0.075 (0.008) & 0.219 (0.034) & 1.122 (0.017) & 1.497 (0.071) \\
& (10, 10$p$, 0.10) & 0.257 (0.051) & 0.098 (0.002) & 0.403 (0.059) & 1.116 (0.043) & 1.410 (0.179) \\
& (10, 10$p$, 0.25) & 0.321 (0.061) & 0.102 (0.003) & 0.465 (0.068) & 1.189 (0.044) & 2.025 (0.352) \\ \cline{2-7}
& (15, 50$p$, 0.10) & 0.068 (0.014) & 0.037 (0.006) & 0.119 (0.018) & 1.903 (0.009) & 1.693 (0.013) \\
& (15, 50$p$, 0.25) & 0.123 (0.021) & 0.064 (0.008) & 0.212 (0.025) & 2.008 (0.015) & 2.009 (0.082) \\
& (15, 10$p$, 0.10) & 0.269 (0.045) & 0.100 (0.002) & 0.434 (0.050) & 1.992 (0.039) & 1.998 (0.227) \\
& (15, 10$p$, 0.25) & 0.349 (0.055) & 0.105 (0.003) & 0.515 (0.059) & 2.074 (0.037) & 3.081 (0.525) \\ \hline
\multirow{12}{*}{\rotatebox[origin=c]{90}{Mod. Friedman function}} 
&  (5, 50$p$, 0.10) & 0.080 (0.019) & 0.093 (0.019) & 0.166 (0.052) & 0.167 (0.013) & 0.409 (0.047) \\
&  (5, 50$p$, 0.25) & 0.118 (0.028) & 0.136 (0.026) & 0.241 (0.070) & 0.225 (0.023) & 0.862 (0.080) \\
&  (5, 10$p$, 0.10) & 0.370 (0.047) & 0.355 (0.004) & 0.913 (0.051) & 0.207 (0.031) & 0.651 (0.168) \\
&  (5, 10$p$, 0.25) & 0.343 (0.049) & 0.352 (0.002) & 0.682 (0.067) & 0.288 (0.058) & 1.220 (0.234) \\ \cline{2-7}
& (10, 50$p$, 0.10) & 0.079 (0.016) & 0.057 (0.014) & 0.134 (0.027) & 0.792 (0.011) & 1.225 (0.019) \\
& (10, 50$p$, 0.25) & 0.128 (0.025) & 0.101 (0.021) & 0.234 (0.042) & 0.864 (0.017) & 1.514 (0.041) \\
& (10, 10$p$, 0.10) & 0.336 (0.035) & 0.350 (0.002) & 0.717 (0.054) & 0.856 (0.039) & 1.480 (0.117) \\
& (10, 10$p$, 0.25) & 0.400 (0.057) & 0.362 (0.003) & 0.931 (0.055) & 0.909 (0.041) & 1.921 (0.261) \\ \cline{2-7}
& (15, 50$p$, 0.10) & 0.072 (0.013) & 0.044 (0.011) & 0.121 (0.020) & 1.444 (0.012) & 1.994 (0.014) \\
& (15, 50$p$, 0.25) & 0.130 (0.020) & 0.084 (0.017) & 0.227 (0.030) & 1.523 (0.016) & 2.179 (0.029) \\
& (15, 10$p$, 0.10) & 0.335 (0.046) & 0.314 (0.017) & 0.853 (0.070) & 1.506 (0.033) & 2.256 (0.168) \\
& (15, 10$p$, 0.25) & 0.404 (0.056) & 0.348 (0.012) & 0.966 (0.063) & 1.562 (0.038) & 3.002 (0.431) \\ \hline
\multirow{12}{*}{\rotatebox[origin=c]{90}{$g-$function}} 
&  (5, 50$p$, 0.10) & 0.374 (0.061) & 0.272 (0.005) & 0.448 (0.069) & 0.320 (0.039) & 1.640 (0.140) \\
&  (5, 50$p$, 0.25) & 0.432 (0.092) & 0.269 (0.005) & 0.530 (0.090) & 0.410 (0.042) & 2.355 (0.185) \\
&  (5, 10$p$, 0.10) & 0.643 (0.108) & 0.280 (0.002) & 0.747 (0.112) & 0.458 (0.103) & 2.217 (0.366) \\
&  (5, 10$p$, 0.25) & 0.697 (0.112) & 0.279 (0.002) & 0.815 (0.116) & 0.539 (0.097) & 2.571 (0.330) \\ \cline{2-7}
& (10, 50$p$, 0.10) & 0.374 (0.057) & 0.326 (0.007) & 0.570 (0.050) & 0.911 (0.022) & 3.407 (0.190) \\
& (10, 50$p$, 0.25) & 0.446 (0.077) & 0.341 (0.007) & 0.707 (0.066) & 1.008 (0.022) & 4.527 (0.207) \\
& (10, 10$p$, 0.10) & 0.688 (0.094) & 0.333 (0.003) & 0.997 (0.092) & 1.052 (0.064) & 4.436 (0.428) \\
& (10, 10$p$, 0.25) & 0.786 (0.105) & 0.336 (0.003) & 1.130 (0.106) & 1.095 (0.053) & 5.039 (0.512) \\ \cline{2-7}
& (15, 50$p$, 0.10) & 0.367 (0.048) & 0.325 (0.012) & 0.584 (0.046) & 1.635 (0.020) & 6.983 (0.267) \\
& (15, 50$p$, 0.25) & 0.444 (0.064) & 0.356 (0.008) & 0.756 (0.052) & 1.527 (0.020) & 5.446 (0.221) \\
& (15, 10$p$, 0.10) & 0.650 (0.089) & 0.348 (0.005) & 1.050 (0.086) & 1.680 (0.052) & 7.474 (0.794) \\
& (15, 10$p$, 0.25) & 0.802 (0.092) & 0.354 (0.005) & 1.240 (0.099) & 1.738 (0.035) & 8.680 (0.815) \\ \hline
\multirow{12}{*}{\rotatebox[origin=c]{90}{Bratley function}} 
&  (5, 50$p$, 0.10) & 0.070 (0.020) & 0.043 (0.007) & 0.099 (0.016) & 0.180 (0.017) & 0.527 (0.039) \\ % 
&  (5, 50$p$, 0.25) & 0.106 (0.037) & 0.050 (0.009) & 0.146 (0.024) & 0.255 (0.028) & 0.962 (0.087) \\ %
&  (5, 10$p$, 0.10) & 0.221 (0.078) & 0.081 (0.002) & 0.276 (0.049) & 0.219 (0.051) & 0.679 (0.100) \\ %
&  (5, 10$p$, 0.25) & 0.293 (0.098) & 0.079 (0.002) & 0.343 (0.068) & 0.288 (0.052) & 1.074 (0.200) \\ \cline{2-7} %
& (10, 50$p$, 0.10) & 0.068 (0.017) & 0.041 (0.007) & 0.122 (0.012) & 1.027 (0.015) & 1.373 (0.032) \\ %
& (10, 50$p$, 0.25) & 0.125 (0.030) & 0.058 (0.009) & 0.200 (0.021) & 1.112 (0.017) & 1.954 (0.087) \\ %
& (10, 10$p$, 0.10) & 0.231 (0.054) & 0.103 (0.001) & 0.376 (0.035) & 1.039 (0.033) & 1.562 (0.145) \\ %
& (10, 10$p$, 0.25) & 0.331 (0.071) & 0.106 (0.002) & 0.472 (0.051) & 1.125 (0.050) & 2.179 (0.283) \\ \cline{2-7} %
& (15, 50$p$, 0.10) & 0.065 (0.015) & 0.033 (0.006) & 0.117 (0.010) & 1.887 (0.009) & 2.234 (0.038) \\ %
& (15, 50$p$, 0.25) & 0.125 (0.021) & 0.054 (0.008) & 0.205 (0.017) & 1.974 (0.017) & 2.990 (0.122) \\ % 
& (15, 10$p$, 0.10) & 0.223 (0.042) & 0.107 (0.002) & 0.394 (0.029) & 1.905 (0.035) & 2.511 (0.166) \\ %
& (15, 10$p$, 0.25) & 0.326 (0.061) & 0.113 (0.002) & 0.504 (0.052) & 1.982 (0.024) & 3.344 (0.253) \\ \hline %
\multirow{12}{*}{\rotatebox[origin=c]{90}{Morris function}} 
&  (5, 50$p$, 0.10) & 0.063 (0.023) & 0.017 (0.003) & 0.064 (0.025) & 0.439 (0.030) & 0.468 (0.110) \\
&  (5, 50$p$, 0.25) & 0.092 (0.035) & 0.018 (0.003) & 0.093 (0.036) & 0.283 (0.028) & 1.238 (0.163) \\
&  (5, 10$p$, 0.10) & 0.137 (0.048) & 0.033 (0.002) & 0.140 (0.049) & 0.463 (0.054) & 0.517 (0.184) \\
&  (5, 10$p$, 0.25) & 0.188 (0.070) & 0.031 (0.003) & 0.191 (0.072) & 0.343 (0.063) & 1.152 (0.315) \\ \cline{2-7}
& (10, 50$p$, 0.10) & 0.079 (0.017) & 0.029 (0.004) & 0.103 (0.016) & 0.627 (0.019) & 0.719 (0.011) \\
& (10, 50$p$, 0.25) & 0.130 (0.024) & 0.044 (0.004) & 0.167 (0.023) & 0.433 (0.024) & 2.072 (0.271) \\
& (10, 10$p$, 0.10) & 0.220 (0.039) & 0.058 (0.002) & 0.284 (0.034) & 0.614 (0.041) & 0.785 (0.114) \\
& (10, 10$p$, 0.25) & 0.295 (0.056) & 0.060 (0.002) & 0.360 (0.051) & 0.457 (0.052) & 1.973 (0.511) \\ \cline{2-7}
& (15, 50$p$, 0.10) & 0.076 (0.013) & 0.026 (0.003) & 0.098 (0.013) & 0.830 (0.018) & 0.958 (0.090) \\
& (15, 50$p$, 0.25) & 0.135 (0.018) & 0.044 (0.005) & 0.180 (0.019) & 0.632 (0.021) & 2.921 (0.353) \\
& (15, 10$p$, 0.10) & 0.239 (0.034) & 0.064 (0.002) & 0.326 (0.031) & 0.805 (0.038) & 0.990 (0.135) \\
& (15, 10$p$, 0.25) & 0.326 (0.043) & 0.069 (0.003) & 0.418 (0.042) & 0.622 (0.043) & 2.775 (0.679) \\ \hline
\end{tabular}
\caption{Estimates of the expected $L_1$ distance between BART-based Sobol\'{} indices and true Sobol\'{} indices when $f(\x)$ is measured with error. 
Each block of four scenarios is ordered roughly in decreasing order of ``signal-to-noise.''}
\label{tbl:raw}
\end{table}

\subsubsection*{Rank-based performance metric}
To answer questions \ref{qu:ranking} and \ref{qu:countrankings}, we replace the $L_1$ distance $d_{L_1}(\bm{\cdot}, \bm{\cdot})$ in Equation \ref{eq:expectation} with a discrepancy measure $d_r(\bm{\cdot}, \bm{\cdot})$, to be defined in Equation \ref{eq:vsum}. This allows a more interpretable comparison between the performances of one-way BART counts and BART-based Sobol' indices. Table \ref{tbl:rnk} shows the sample mean and standard deviation of these 500 estimates.

\begin{table}
\tiny
\centering
% \begin{tabular}{|*5{c|}}
\begin{tabular}{|c | r || *5{l|}}
\cline{2-7}
\multicolumn{1}{c|}{\multirow{1}{*}{}} & 
 ($p$, $n$, $\sigma^2$) &
 $S_i^{\ens}$ vs $S_i^f$ & 
 Count vs $S_i^f$ & 
 $T_i^{\ens}$ vs $T_i^f$ & 
 Count vs $T_i^f$ & 
 $S_{ij}^{\ens}$ vs $S_{ij}^f$ \\  \cline{2-7}  \cline{2-7} \hline\hline
\multirow{15}{*}{\rotatebox[origin=c]{90}{Friedman function}} 
& Max value & 20 & 20 & 20 & 20 & 20  \\
& (5, 50$p$, 0.10) & 1.048 (1.000) & 5.420 (2.195) & 1.040 (1.000) & 5.420 (2.195) & 0.000 (0.000) \\
& (5, 50$p$, 0.25) & 1.152 (1.005) & 6.404 (3.186) & 1.144 (1.015) & 6.404 (3.186) & 0.008 (0.126)  \\
& (5, 10$p$, 0.10) & 2.540 (1.484) & 8.620 (3.011) & 2.536 (1.488) & 8.620 (3.011) & 8.716 (5.690) \\
& (5, 10$p$, 0.25) & 2.816 (1.892) & 8.916 (3.086) & 2.812 (1.888) & 8.916 (3.086) & 9.848 (5.696) \\ \cline{2-7}
& Max value & 70 & 70 & 70 & 70 & 90  \\
& (10, 50$p$, 0.10) & 0.976 (1.001) & 4.984 (1.430) & 0.960 (1.000) & 4.984 (1.430) & 0.000 (0.000) \\
& (10, 50$p$, 0.25) & 1.136 (0.992) & 6.144 (2.776) & 1.108 (0.995) & 6.144 (2.776) & 0.000 (0.000) \\
& (10, 10$p$, 0.10) & 2.020 (0.900) & 18.608 (8.795) & 2.016 (0.904) & 18.608 (8.795) & 4.820 (9.915) \\
& (10, 10$p$, 0.25) & 2.412 (1.591) & 23.844 (10.447) & 2.424 (1.615) & 23.844 (10.447) & 12.772 (16.712) \\ \cline{2-7}
& Max value & 120 & 120 & 120 & 120 & 210  \\
& (15, 50$p$, 0.10) & 0.792 (0.979) & 4.432 (0.862) & 0.768 (0.974) & 4.432 (0.862) & 0.000 (0.000) \\
& (15, 50$p$, 0.25) & 1.008 (1.001) & 5.172 (1.919) & 1.008 (1.001) & 5.172 (1.919) & 0.000 (0.000) \\
& (15, 10$p$, 0.10) & 1.780 (0.754) & 8.504 (5.658) & 1.772 (0.762) & 8.504 (5.658) & 0.376 (1.647) \\
& (15, 10$p$, 0.25) & 1.988 (1.093) & 15.456 (10.630) & 1.988 (1.107) & 15.456 (10.630) & 3.760 (9.419) \\ \hline 
\multirow{15}{*}{\rotatebox[origin=c]{90}{Mod. Friedman function}} 
& Max value & 20 & 20 & 20 & 20 & 20  \\
& (5, 50$p$, 0.10) & 1.024 (1.001) & 13.292 (1.359) & 1.004 (1.001) & 5.300 (1.363) & 0.000 (0.000)\\
& (5, 50$p$, 0.25) & 1.004 (1.001) & 13.728 (1.761) & 1.024 (1.001) & 5.752 (1.791) & 0.000 (0.000) \\
& (5, 10$p$, 0.10) & 1.864 (0.789) & 6.796 (5.295) & 13.896 (4.901) & 27.528 (8.823) & 0.000 (0.000) \\
& (5, 10$p$, 0.25) & 2.020 (1.211) & 3.960 (2.851) & 9.280 (1.257) & 9.360 (2.678) & 0.136 (0.656) \\ \cline{2-7}
& Max value & 70 & 70 & 70 & 70 & 90  \\
& (10, 50$p$, 0.10) & 0.856 (0.991) & 12.736 (1.098) & 0.860 (0.991) & 4.736 (1.098) & 0.000 (0.000)  \\
& (10, 50$p$, 0.25) & 1.024 (1.001) & 13.940 (2.078) & 1.020 (1.001) & 5.940 (2.078) & 0.000 (0.000) \\
& (10, 10$p$, 0.10) & 1.888 (0.832) & 3.840 (2.521) & 9.648 (0.842) & 9.392 (2.665) & 0.000 (0.000) \\
& (10, 10$p$, 0.25) & 1.912 (1.390) & 8.552 (5.685) & 17.740 (5.670) & 31.520 (8.478) & 0.000 (0.000) \\ \cline{2-7}
& Max value & 120 & 120 & 120 & 120 & 210  \\
& (15, 50$p$, 0.10) & 0.696 (0.954) & 12.500 (0.867) & 0.680 (0.948) & 4.500 (0.867) & 0.000 (0.000) \\
& (15, 50$p$, 0.25) & 0.976 (1.001) & 12.940 (1.348) & 0.952 (1.000) & 4.948 (1.354) & 0.000 (0.000) \\
& (15, 10$p$, 0.10) & 1.724 (0.690) & 5.292 (4.161) & 8.096 (2.377) & 12.596 (10.107) & 0.000 (0.000) \\
& (15, 10$p$, 0.25) & 1.792 (0.958) & 5.716 (5.927) & 10.928 (4.388) & 29.084 (15.196) & 0.000 (0.000) \\ \hline 
\multirow{15}{*}{\rotatebox[origin=c]{90}{$g-$function}} 
& Max value & 20 & 20 & 20 & 20 & 90  \\
& (5, 50$p$, 0.10) & 1.068 (1.311) & 5.768 (3.576) & 1.092 (1.328) & 5.768 (3.576) & 35.640 (10.136) \\
& (5, 50$p$, 0.25) & 1.920 (1.757) & 8.492 (4.441) & 1.968 (1.759) & 8.492 (4.441) & 42.168 (12.003) \\
& (5, 10$p$, 0.10) & 4.636 (3.029) & 12.864 (4.367) & 4.648 (3.020) & 12.864 (4.367) & 52.236 (13.342) \\
& (5, 10$p$, 0.25) & 6.160 (3.358) & 13.576 (4.197) & 6.204 (3.376) & 13.576 (4.197) & 54.632 (13.402) \\ \cline{2-7}
& Max value & 70 & 70 & 70 & 70 & 790  \\
& (10, 50$p$, 0.10) & 0.652 (0.996) & 4.188 (3.201) & 0.640 (0.984) & 4.188 (3.201) & 175.484 (52.348) \\
& (10, 50$p$, 0.25) & 1.392 (1.591) & 8.816 (5.697) & 1.444 (1.647) & 8.816 (5.697) & 252.916 (64.184) \\
& (10, 10$p$, 0.10) & 6.388 (4.313) & 27.040 (12.884) & 6.428 (4.286) & 27.040 (12.884) & 338.148 (87.447) \\
& (10, 10$p$, 0.25) & 10.500 (6.133) & 35.024 (12.362) & 10.668 (6.162) & 35.024 (12.362) & 377.112 (88.710) \\ \cline{2-7}
& Max value & 120 & 120 & 120 & 120 & 1990  \\
& (15, 50$p$, 0.10) & 0.408 (0.817) & 2.088 (2.360) & 0.400 (0.811) & 2.088 (2.360) & 260.228 (98.276) \\
& (15, 50$p$, 0.25) & 0.864 (1.177) & 6.120 (4.861) & 0.928 (1.209) & 6.120 (4.861) & 456.692 (133.275) \\
& (15, 10$p$, 0.10) & 4.780 (4.517) & 18.124 (10.946) & 4.808 (4.544) & 18.124 (10.946) & 578.464 (172.922) \\
& (15, 10$p$, 0.25) & 10.908 (7.537) & 34.672 (18.045) & 11.016 (7.569) & 34.672 (18.045) & 733.748 (211.209) \\ \hline
\multirow{15}{*}{\rotatebox[origin=c]{90}{Bratley function}} 
& Max value & 20 & 20 & 20 & 20 & 90  \\
& (5, 50$p$, 0.10) & 0.000 (0.000) & 0.656 (1.097) & 0.000 (0.000) & 0.656 (1.097) & 0.124 (0.627) \\ %
& (5, 50$p$, 0.25) & 0.008 (0.126) & 2.004 (2.188) & 0.000 (0.000) & 2.004 (2.188) & 1.056 (1.991) \\ %
& (5, 10$p$, 0.10) & 0.280 (0.771) & 9.132 (4.824) & 0.280 (0.771) & 9.132 (4.824) & 11.872 (8.695) \\ %
& (5, 10$p$, 0.25) & 0.716 (1.231) & 9.932 (4.746) & 0.716 (1.237) & 9.932 (4.746) & 15.480 (9.515) \\ \cline{2-7} % 
& Max value & 70 & 70 & 70 & 70 & 790  \\
& (10, 50$p$, 0.10) & 0.084 (0.553) & 2.124 (2.933) & 0.004 (0.089) & 2.124 (2.933) & 28.348 (26.519) \\ %
& (10, 50$p$, 0.25) & 1.236 (2.459) & 4.884 (4.212) & 0.580 (1.495) & 4.884 (4.212) & 79.328 (45.473) \\ %
& (10, 10$p$, 0.10) & 5.352 (4.851) & 20.588 (8.920) & 5.280 (4.784) & 20.588 (8.920) & 197.556 (60.194) \\ %
& (10, 10$p$, 0.25) & 7.352 (4.991) & 25.356 (9.913) & 7.284 (4.890) & 25.356 (9.913) & 237.864 (69.558) \\ \cline{2-7} %
& Max value & 120 & 120 & 120 & 120 & 1990  \\
& (15, 50$p$, 0.10) & 0.020 (0.236) & 0.948 (1.905) & 0.000 (0.000) & 0.948 (1.905) & 22.344 (31.235) \\ %
& (15, 50$p$, 0.25) & 1.216 (2.566) & 6.600 (6.780) & 0.464 (1.517) & 6.600 (6.780) & 126.056 (91.807) \\
& (15, 10$p$, 0.10) & 7.052 (7.614) & 24.312 (12.077) & 7.008 (7.603) & 24.312 (12.077) & 402.292 (128.257)  \\ %
& (15, 10$p$, 0.25) & 11.728 (9.074) & 30.804 (12.989) & 11.636 (9.031) & 30.804 (12.989) & 483.504 (146.360) \\ \hline %
\multirow{15}{*}{\rotatebox[origin=c]{90}{Morris function}} 
& Max value & 20 & 20 & 20 & 20 & 90  \\
& (5, 50$p$, 0.10) & 0.000 (0.000) & 0.000 (0.000) & 0.000 (0.000) & 0.000 (0.000) & 0.000 (0.000) \\
& (5, 50$p$, 0.25) & 0.000 (0.000) & 0.000 (0.000) & 0.000 (0.000) & 0.000 (0.000) & 0.000 (0.000) \\
& (5, 10$p$, 0.10) & 0.000 (0.000) & 0.000 (0.000) & 0.000 (0.000) & 0.000 (0.000) & 0.000 (0.000) \\
& (5, 10$p$, 0.25) & 0.000 (0.000) & 0.000 (0.000) & 0.000 (0.000) & 0.000 (0.000) & 0.000 (0.000) \\ \cline{2-7}
& Max value & 70 & 70 & 70 & 70 & 790  \\
& (10, 50$p$, 0.10) & 0.000 (0.000) & 0.000 (0.000) & 0.000 (0.000) & 0.000 (0.000) & 1.396 (3.844) \\
& (10, 50$p$, 0.25) & 0.000 (0.000) & 0.240 (0.951) & 0.000 (0.000) & 0.240 (0.951) & 25.252 (24.050) \\
& (10, 10$p$, 0.10) & 0.000 (0.000) & 11.880 (7.979) & 0.000 (0.000) & 11.880 (7.979) & 129.608 (65.530) \\
& (10, 10$p$, 0.25) & 0.000 (0.000) & 15.144 (9.359) & 0.000 (0.000) & 15.144 (9.359) & 191.964 (78.236) \\ \cline{2-7}
& Max value & 120 & 120 & 120 & 120 & 1990  \\
& (15, 50$p$, 0.10) & 0.000 (0.000) & 0.000 (0.000) & 0.000 (0.000) & 0.000 (0.000) & 0.304 (2.605) \\
& (15, 50$p$, 0.25) & 0.000 (0.000) & 0.008 (0.126) & 0.000 (0.000) & 0.008 (0.126) & 19.228 (24.255) \\
& (15, 10$p$, 0.10) & 0.000 (0.000) & 2.800 (4.816) & 0.000 (0.000) & 2.800 (4.816) & 137.684 (91.270) \\
& (15, 10$p$, 0.25) & 0.000 (0.000) & 8.124 (8.533) & 0.000 (0.000) & 8.124 (8.533) & 269.444 (125.489) \\ \hline
\end{tabular}
\caption{Estimates of the expected $d_r$ discrepancy between BART-based Sobol\'{} index rankings and true Sobol\'{} index rankings when $f(\x)$ is measured with error.  
Each block of four scenarios is ordered roughly in decreasing order of ``signal-to-noise.''}
\label{tbl:rnk}
\end{table}

As an example, we will rank the Friedman function's normalized first-order Sobol\'{} index values $(S_1, S_2, S_3, S_4, S_5) = (.197, .197, .093, .350, .087)$ shown in Table \ref{tbl:VA1} as $(2, 2, 4, 1, 5)$, where the most active variable (i.e. variable 4) is assigned ranking number 1 and the least active variable (i.e. variable 5) is assigned ranking number 5. Variables 1 and 2 are equally active, so we will adopt the convention used in many sports competitions of assigning the minimum rank to the two variables and then leaving a gap in the ranking numbers so that the positions of all variables less active than variables 1 and 2 are unaffected. 

Several options exist for comparing two rankings. \cite{Kendall48} introduces a distance that, when ties in rankings are not allowed, is the graphical distance between two vertices in the well-studied permutation polytope that represents all possible rankings of $p$ objects \citep{Heiser13}. \cite{Emond2002} point out that when ties are allowed, Kendall's ``distance'' violates the triangle inequality and hence is no longer a true metric. They advocate the distance defined by \cite{Kemeny1962}, which equals Kendall's distance when ties are not allowed, but remains a metric when ties are allowed. 

Unfortunately, the Kemeny-Snell (KS) distance is likely to artificially inflate when the data-generating function has either more than two inert variables or has equally-active non-inert variables. In the former scenario, a fitted BART model is unlikely to entirely shrink all of its input activity measures of the inert variables. In this case, the KS distance will be inflated by the fitted BART model assigning small but positive effects to the inert variables. In the latter scenario, a fitted BART model is unlikely to perfectly match its input activity measures of the equally-active non-inert variables. In this case, the fitted BART model could be incorrectly ``punished'' for even the slightest discrepancy between the variable-activity measures of two equally-active variables. To resolve this issue, we create a discrepancy measure based on the multi-stage discordance measures discussed in \cite{Fligner1988}.

To compute the discordance measures between two rankings $\rho_f$ and $\rho_{\ens}$, where in our variable activity setting $\rho_f$ represents the true input activity and $\rho_{\ens}$ represents the input activity of our fitted BART model, \cite{Fligner1988} assume that neither ranking has any ties.
As an example, suppose $\rho_f = (4, 3, 1, 2)$ and $\rho_{\ens} = (3, 1, 2, 4)$. 
The discordances $W_1, W_2, \ldots, W_4$ will be computed sequentially. To compute discordance $W_1$, we see that variable $3$ is the most active in $\rho_f$. Since variable $3$ is the second most active in $\rho_{\ens}$, we set $W_1 = 2 - 1 = 1$. We then remove variable $3$ from consideration to compute $W_2$, $W_3$, and $W_4$. To compute discordance $W_2$, we see that variable $4$ is the most active of the remaining variables ($1$, $2$, and $4$) in $\rho_f$. Since variable $4$ is the third most active of the remaining variables ($1$, $2$, and $4$) in $\rho_{\ens}$, we set $W_2 = 3 - 1 = 2$. We then remove variable $4$ from consideration to compute discordances $W_3$ and $W_4$. To compute $W_3$, we see that variable $2$ is the most active of the remaining variables ($1$ and $2$) in $\rho_f$. Since variable $2$ is the most active of the remaining variables ($1$ and $2$) in $\rho_{\ens}$, we set $W_3 = 1 - 1 = 0$. We then remove variable $2$ from consideration to compute $W_4$. Since only one variable remains, we set $W_4 = 0$. Hence, the discordances in this example are $(W_1, W_2, W_3, W_4) = (1, 2, 0, 0)$.

More generally (but still assuming neither ranking has any ties), suppose we have already computed discordances $W_1, \ldots, W_{k-1}$ for some $k = 1, \ldots, q$, where $q$ is the number of elements in vector $\rho_f$, and wish to compute $W_k$. Note that $q$ is not necessarily $p$ (e.g. $q = \binom{p}{2}$ when the rankings represent two-way interactions). Thus, we have removed $k-1$ of the $q$ items (e.g. variables, variable pairs, variable triplets) from consideration. If item $i$ is the most active in ranking $\rho_f$ among the remaining considered items, we then find $j$, where item $i$ is the $j$th most active in ranking $\rho_{\ens}$ among the remaining considered items. The value $W_k$ is then set to be $j - 1$. 

Now suppose both ranking $\rho_f$ and ranking $\rho_{\ens}$ are allowed to have ties. As mentioned earlier, we will adopt the ``standard competition'' ranking convention of, for each set of items tied with each other, assigning the minimum rank to the tied items and then leaving a gap in the ranking numbers so that the positions of all items less active than the tied items are unaffected. For example, first-order Sobol\'{} index values $(0.1, 0.1, 0.2, 0.2, 0.35, 0.05)$ would be ranked $(4, 4, 2, 2, 1, 6)$.
Suppose we have already computed $W_1, \ldots, W_{k-1}$ for some $k = 1, \ldots, q$ and wish to compute $W_k$. If $u \geq 1$ items $i_1, \ldots, i_u$ are tied for most active in ranking $\rho_f$ among the remaining considered items, we can find $j_1, \ldots, j_u$, where item $i_l$ is the $j_l$th most active item in ranking $\rho_{\ens}$ among the remaining considered items. The value $W_k$ is then set to be $\min_{l = 1, \ldots, u} j_l - 1$. If $argmin_{l = 1, \ldots, u} j_l$ has more than one value, then we pick any one (it does not matter which) of the corresponding items $i_l$ to remove from consideration. Once an item is removed from consideration, the value $W_{k+1}$ can then be computed (if $k < p$). Note that if $u = 1$, this reduces to the ``no-ties'' case. 

We can now define our discrepancy measure between rankings $\rho_f$ and $\rho_{\ens}$:
\begin{equation}
d_r(\rho_f, \rho_{\ens}) = 2 \sum_{k=1}^{q} W_k,
\label{eq:vsum}
\end{equation}
where discordances $W_1, W_2, \ldots, W_q$ are computed as described in the previous paragraph. This measure has three particularly desirable properties. First, it equals Kendall's distance (and hence the KS distance) when ties are not allowed. Second, it does not inflate as the number of data-generating function $f$'s inert variables increases. In particular, discordance $W_k = 0$ for all $k > q_0$, where $q_0$ is the number of active items (i.e. items with non-zero input activity measure) in $f$. Hence, the discrepancy measure is invariant to the number of inert variables. Third, it does not inflate when $f$ has equally-active non-inert items. If $f$ has a set of equally-active non-inert items, then the discrepancy measure will not inflate as long as the equally-active items in the set are consecutively ranked. These three properties can be stated as Theorems \ref{thm:dist}, \ref{thm:inert}, and \ref{thm:tied} whose proofs are in the appendix.

\begin{thm}
If rankings $\bm{\alpha}$ and $\bm{\beta}$ each have no ties, then the Kemeny-Snell distance between $\bm{\alpha}$ and $\bm{\beta}$ equals the discrepancy measure $d_r(\bm{\alpha}, \bm{\beta})$ in Equation \ref{eq:vsum}.
\label{thm:dist}
\end{thm}

\begin{thm}
Consider the discrepancy $d_r(\rho_f, \rho_{\ens})$ between rankings $\rho_f$ and $\rho_{\ens}$. Then the discordance $W_k = 0$ for all $k > q_0$, where $q_0$ is the number of active items in $f$.
\label{thm:inert}
\end{thm}

\begin{thm}
Consider the discrepancy $d_r(\rho_f, \rho_{\ens})$ between rankings $\rho_f$ and $\rho_{\ens}$. Suppose $u \geq 1$ items $i_{j}, \ldots, i_{j+u-1}$ (and no other items) have ranking number $j$ in $\rho_f$ and ranking numbers $r_{j}, \ldots, r_{j+u-1}$ in $\rho_{\ens}$. 
Then all $|\rho_f|$ discordances are invariant to choice of permutation $\phi$ of set $\{r_{j}, \ldots, r_{j+u-1}\}$ of ranking numbers. 
\label{thm:tied}
\end{thm}

\subsection{Simulation results}
\label{sec:simres}

To answer question \ref{qu:accuracy2}, we find in Table \ref{tbl:raw} that when the underlying data-generating function is not known, our BART-based Sobol\'{} indices should always be preferred over TGP-based Sobol\'{} indices. For all five data-generating functions, and for $p = 10$ or $15$, if we compare the BART and TGP ``$S_i$'' columns to each other and the BART and TGP ``$T_i$'' columns to each other, the BART-based $L_1$ distances are uniformly lower than the TGP-based $L_1$ distances, often by a dramatic amount. For $p=5$, this observation does not hold uniformly (e.g. for the $(\text{function}, p, n, \sigma^2) = (\text{Mod. Friedman}, 5, 10p, 0.10)$ row, the TGP-based total-effects $L_1$ distance is lower than that of BART), but is still valid in most of these cases. 

To answer question \ref{qu:accuracy}, we find in the BART columns of Table \ref{tbl:raw} that the bias of BART-based Sobol\'{} indices is largest with the multiplicative $g-$function and increases with increasing noise and decreasing sample size.

For the five data-generating functions, the $L_1$ distances for all five data-generating functions and all three Sobol\'{} index measures tend to increase with each of increasing noise and decreasing sample size. That is, for a given number $p$ of variables, the BART-based Sobol\'{} indices perform better as ``signal-to-noise'' ratio increases. 

Interestingly, for the Friedman, Modified Friedman, $g-$, and Bratley functions, and for each set of four scenarios, the performance difference between data scenarios $(n = 50p, \sigma^2 = 0.10\Var(f(\X)))$ and $(n = 50p, \sigma^2 = 0.25\Var(f(\X)))$ is much smaller than the performance differences between $(50p, 0.25)$ and $(10p, 0.10)$ and between $(10p, 0.10)$ and $(10p, 0.25)$. We might infer that the $n = 50p$ scenario saturates the data with enough signal for modest noise increases to not make much of a performance difference, but the change from $50p$ to $10p$ makes the signal so scarce that modest noise increases does make a performance difference. These differences seem to be more evenly spread out for the Morris function. 

Recall from Table \ref{tbl:VA1} that for the modified Friedman function, first-order indices $S_1^f = S_2^f = 0$ while total-effects indices $T_1^f = T_2^f = 0.335$. That is, variables $x_1$ and $x_2$ interact strongly with other inputs but are not important on their own. We see in Table \ref{tbl:raw} that in the $(p = 5, n = 50p, \sigma^2 = 0.10\Var(f(\X)))$ and $(p = 5, n = 50p, \sigma^2 = 0.25\Var(f(\X)))$ scenarios, BART captures both the first-order and total-effects indices of the modified Friedman function about as well as it captures the same indices of the original Friedman function. This implies that with enough signal, BART is able to tell if an input is important on its own or if it merely interacts strongly with other inputs. 

Also perhaps unsurprisingly, BART performs worse with the multiplicative $g-$function than it does with the four other data-generating functions. The Friedman function, modified Friedman function, and Morris function each is a sum of either univariate or bivariate functions which BART's additive structure can presumably capture well. The Bratley function is a sum of five simple terms with two of them being either univariate or bivariate. On the other hand, our $g-$function is a product of five univariate functions. If we note that the log of our $g-$function is also a sum of univariate functions, we might expect BART to perform better if we took the log of the $g-$function response data. 

Finally, BART tends to capture total-effects indices less accurately than it does first-order indices. Interestingly, this observation holds even for the high-signal scenarios. 
To answer questions \ref{qu:ranking} and \ref{qu:countrankings}, we find in Table \ref{tbl:rnk} that the first-order rankings provided by BART-based Sobol\'{} indices are uniformly more accurate than those provided by one-way BART counts. First, the ``$S_i^{\ens}$ vs $S_i^f$'' column implies our BART-based first-order Sobol\'{} indices perform incredibly well at predicting the correct order of the true first-order Sobol\'{} indices across all data scenarios for the original Friedman, modified Friedman, and Morris functions, and across the $n = 50p$ scenarios for the $g-$ and Bratley functions. For all Morris-function scenarios, all $500$ sets of BART-based first-order Sobol\'{} indices correctly rank the first $p_0=5$ input variables as more active than any of the other $p-p_0$ input variables.

Second, the ``$T_i^{\ens}$ vs $T_i^f$'' column implies our BART-based total-effects Sobol\'{} indices also perform very well at predicting the correct order of the true total-effects Sobol\'{} indices across all data scenarios for the original Friedman and Morris functions, and across the $n = 50p$ scenarios for the modified Friedman, $g-$, and Bratley functions. Again, for all Morris-function scenarios, all $500$ sets of BART-based total-effects Sobol\'{} indices correctly rank the first $p_0=5$ input variables as more active than any of the other $p-p_0$ input variables.

Finally, for each row, the ``$S_i^{\ens}$ vs $S_i^f$'' expected discrepancy estimate is the same\footnote{We say two estimates $est_1 (se_1)$ and $est_2 (se_2)$ are the same if the intervals produced by the estimate plus or minus the shown standard error overlap (i.e. if interval $(est_1 - se_1, est_1 + se_1)$ and interval $(est_2 - se_2, est_2 + se_2)$ overlap).} or lower than the ``Count vs $S_i^f$'' expected discrepancy estimate. Similarly, the ``$T_i^{\ens}$ vs $T_i^f$'' expected discrepancy estimate is the same or lower than the ``Count vs $T_i^f$'' expected discrepancy estimate for all rows. These observations imply that our BART-based Sobol\'{} indices outperform one-way BART counts across the board when predicting the correct order of the true first-order Sobol\'{} indices and of the true total-effects Sobol\'{} indices. Hence, our BART-based first-order and total-effects Sobol\'{} indices should always be preferred over one-way counts. 

We conclude that our BART-based first-order Sobol\'{} indices can accurately predict the raw values of first-order Sobol\'{} indices of additive data-generating functions. Also, our BART-based first-order and total-effects Sobol\'{} indices can accurately predict the rankings of, respectively, first-order Sobol\'{} indices and total-effects Sobol\'{} indices of both additive and multiplicative data-generating functions. Finally, our BART-based first-order and total-effects Sobol\'{} indices outperform one-way BART counts for all three data-generating functions.

\subsection{Application to the En-ROADS Climate Simulator}

We compute Sobol\'{} indices for a BART model trained on data generated from the En-ROADS climate simulator \citep{enroads2}. This simulator is a mathematical model of how global temperature and carbon emissions, among other factors, are influenced by changes in energy, land use, consumption, agriculture, and other policies. It is designed to be easily used by policymakers, educators, and the general public. The model, an ordinary differential equation solved by Euler integration, synthesizes what its developers consider to be the best available climate science. This simulator is available from the Climate Interactive website.

\begin{figure}[ht!]
    \centering
    \includegraphics{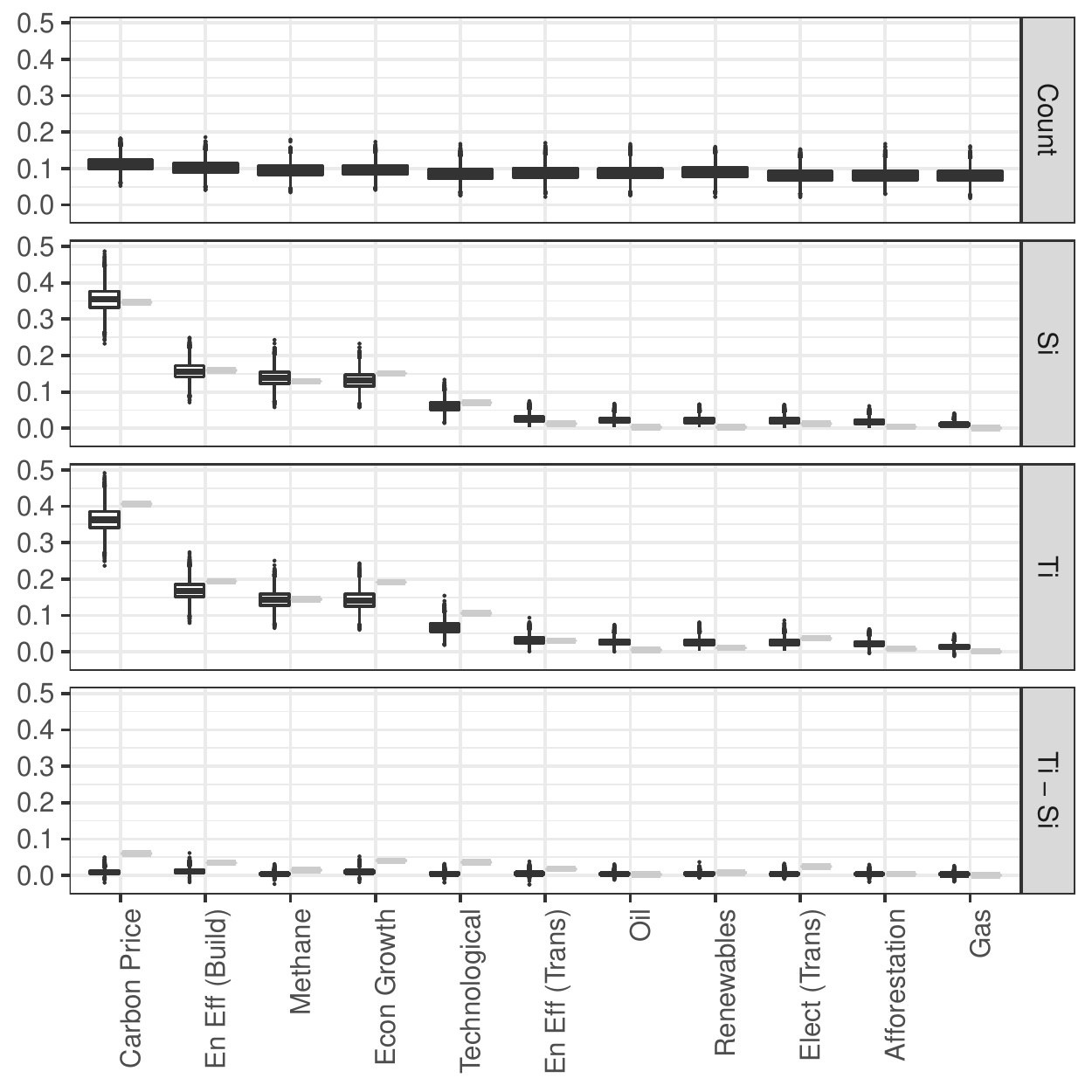}
    \caption{Variable activity measures of BART and GP models trained on data from En-ROADS climate simulator. Variable counts (top panel), BART-based first-order Sobol\'{} indices (second panel), BART-based total-effects Sobol\'{} indices (third panel), and the difference between total-effects and first-order (bottom panel) are shown. Variable activity measures of the 10,000 ensembles corresponding to posterior samples of the trained BART model are shown in black. Point estimates of Sobol\'{} indices of the trained GP model based on the same data are shown in grey.}
    \label{fig:climsobol}
\end{figure}

For this paper, we looked specifically at how the average global temperature increase by 2100 from pre-industrial levels is influenced by the 18 ``top-level'' input variables shown when the En-ROADS climate simulator is first loaded on to a web browser. We explored a subset of 11 variables as summarized in Fig \ref{fig:climsobol} and left the remaining 7 variables at their default settings. Based on an initial exploratory analysis, we found these 7 variables to be either redundant (coal, bioenergy, nuclear, electrification of buildings and industry, deforestation), impractical or unethical to control (population growth), or too discrete to treat as a continuous variable (technological carbon removal). Each input variable is bounded by a minimum and maximum value. We found a maximin LHD of $10 \times 11 = 110$ points on $[0, 1]^{11}$ and scaled it so that the design space contained the range of possible values. However, the simulator rounds values entered into its text fields, effectively rounding each design point to the nearest point on the induced 11-dimensional grid. We then manually obtained response values for each design point. The simulator also rounded the response values to the nearest first decimal place. Because a $0.1^{\circ} F$ difference is smaller than a $0.1^{\circ} C$ difference, we used Fahrenheit values. We then rescaled this ``rounded'' maximin LHD design back onto $[0, 1]^{11}$ to which we trained a BART model with the default parameter settings of, in particular, 10,000 posterior samples from the distribution in Equation \ref{eq:BARTposterior} and 200 trees. For this climate application, we use the \texttt{BART} R package \citep{BART19}.

We computed the first-order, second-order, and total-effects Sobol\'{} indices of the BART model trained on our collected climate simulator data. Because main effects account for more than $96\%$ of the BART model's total variance, we do not show two-way Sobol\'{} indices. By taking the mean first-order Sobol\'{} indices of the 11 input variables over the 10,000 posterior samples, we see in Figure \ref{fig:climsobol} that carbon price accounts for $35.1\%$ of the BART model's total variance, which is twice as much as the next largest impacts of energy efficiency of buildings and industry at $16.1\%$, methane and other (which includes nitrous oxide and fluorinated gases) at $14.4\%$, and economic growth at $13.1\%$. The total-effects Sobol\'{} indices imply a similar conclusion. Variable counts fail to provide evidence of such large differences in impacts of the input variables. 

\begin{figure}[ht!]
    \centering
    \begin{subfigure}[t]{0.48\textwidth}
        \centering
        \includegraphics[width=0.95\textwidth]{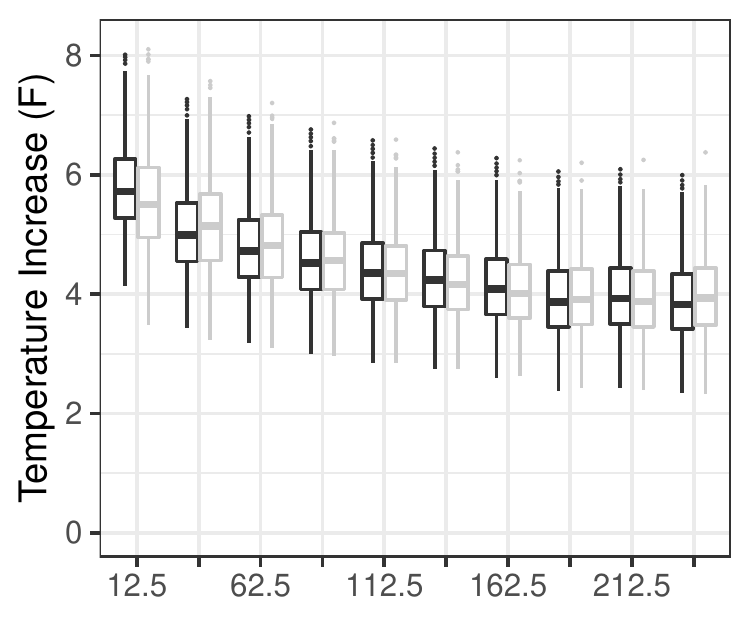}
        \caption{Carbon Price}
        \label{fig:climrange4}
    \end{subfigure}
    ~
    \begin{subfigure}[t]{0.48\textwidth}
        \centering
        \includegraphics[width=0.95\textwidth]{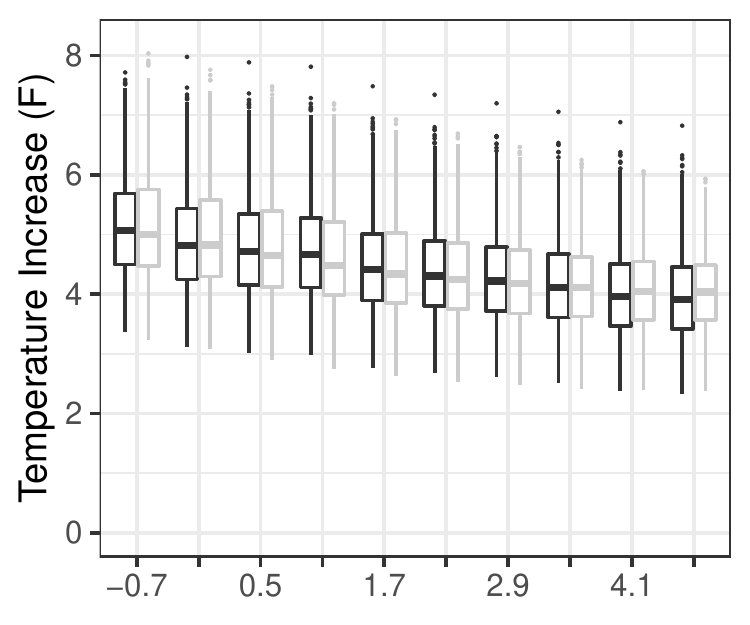}
        \caption{Energy Efficiency (Buildings and Industry)}
        \label{fig:climrange7}
    \end{subfigure}
    ~
    \begin{subfigure}[t]{0.48\textwidth}
        \centering
        \includegraphics[width=0.95\textwidth]{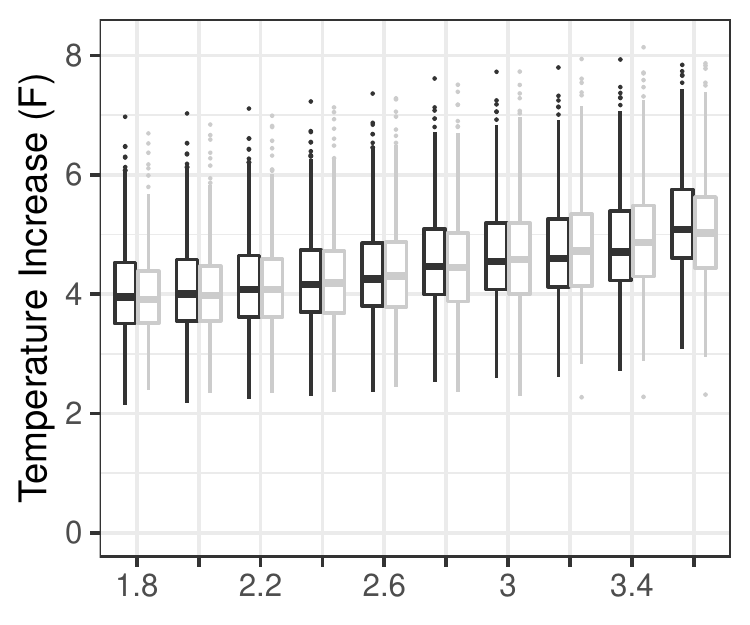}
        \caption{Economic Growth}
        \label{fig:climrange8}
    \end{subfigure}
    ~
    \begin{subfigure}[t]{0.48\textwidth}
        \centering
        \includegraphics[width=0.95\textwidth]{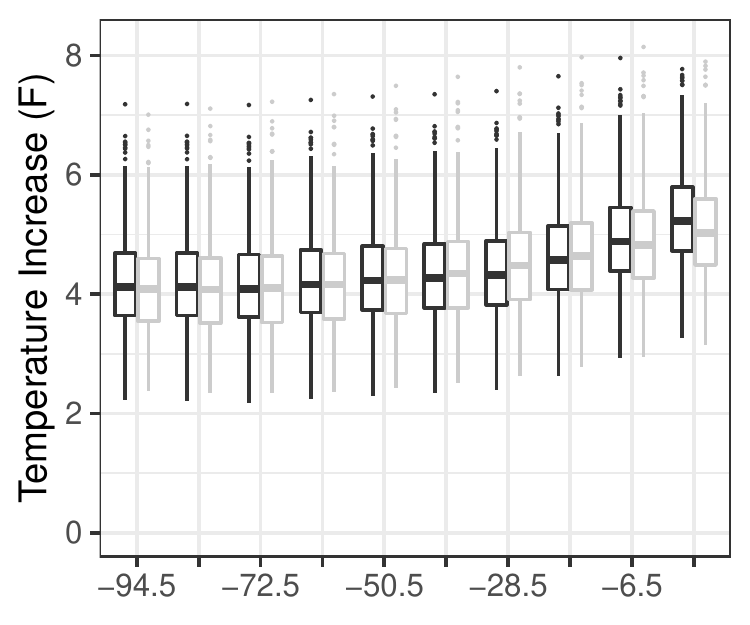}
        \caption{Methane \& Other}
        \label{fig:climrange9}
    \end{subfigure}
    \caption{Range plots of the four most active input variables. The trained BART model is shown in black while the trained GP model is shown in grey.}
    \label{fig:climmaineffects}
\end{figure}

For comparison, we also estimated sensitivity indices and created range plots \citep[see][Chapter 7]{Santner18} by fitting a constant mean Gaussian process (GP) model \citep[i.e. kriging model; see][]{Cressie2015} to the training data. In Figure \ref{fig:climsobol}, we see that the Sobol\'{} indices of the trained BART model match those of the trained kriging model quite well. We also see that the one-way counts of the trained BART model poorly matches both the first-order and total-effects Sobol\'{} indices of the trained kriging model, which supports the hypothesis that the variable count heuristic is not a meaningful input activity measure. In Figure \ref{fig:climmaineffects}, we show range plots of the four most active input variables. For each input, we approximated the marginal response at each of 10 equally spaced points by varying the remaining inputs using a $2^9$-point Sobol\'{} sequence design \citep{SobolSequence}. In the left two plots (carbon price and energy efficiency of buildings and industry), the response and its slope decrease with increasing input values. In the right two plots (economic growth and methane \& other), the response and its slope increase with increasing input values. Hence, all four of these input variables seem to marginally have diminishing effects on future temperature increase.

\begin{figure}[ht!]
    \centering
    \begin{subfigure}[t]{0.48\textwidth}
        \centering
        \includegraphics[width=0.95\textwidth]{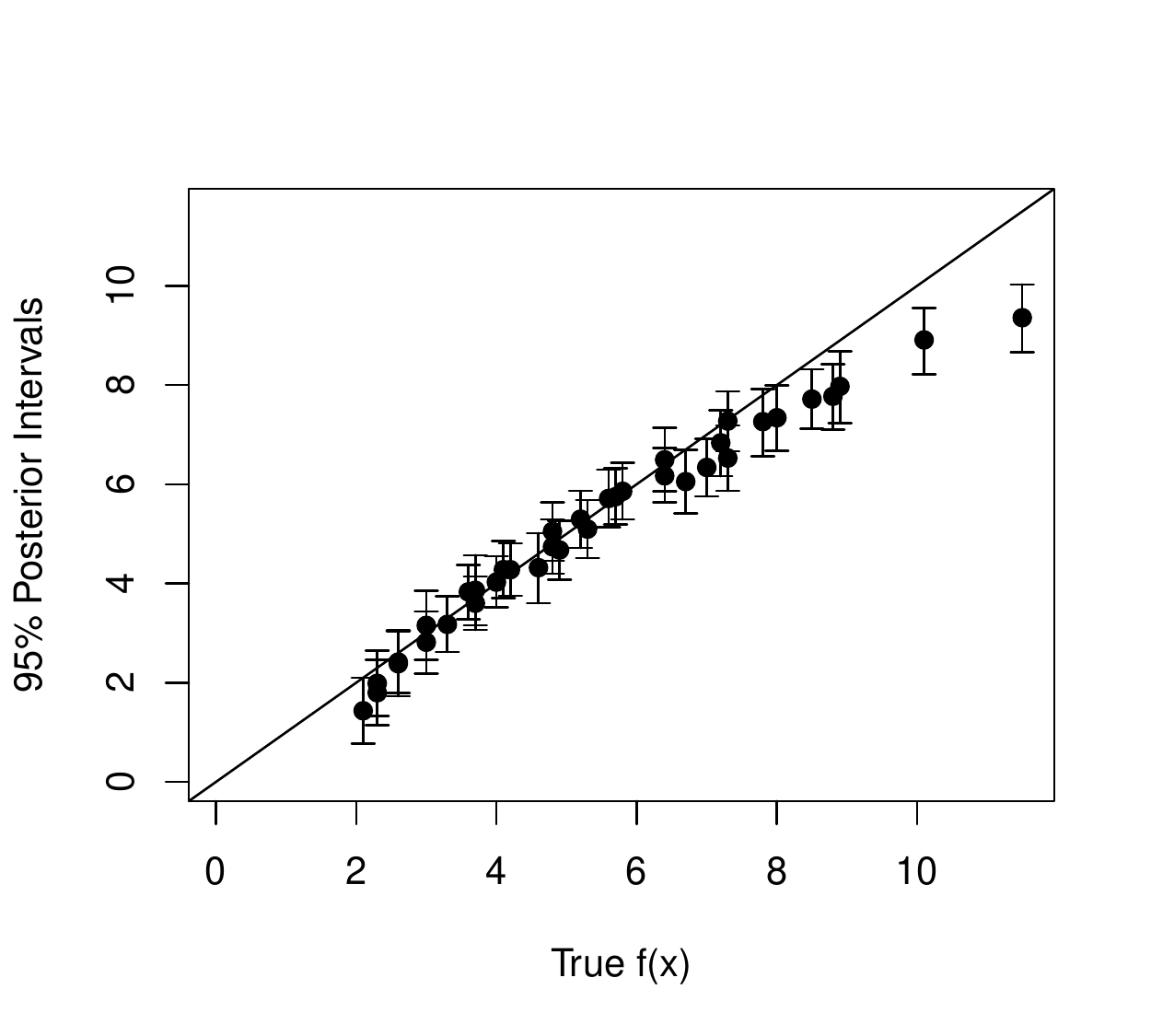}
        \caption{Posterior prediction intervals (PPIs) for 37 out-of-sample test values of $f(\x)$. For each test input point $\x$, error bars reflect the 0.025 and 0.975 quantiles of the 10,000 posterior predictions of $f(\x)$.}
        \label{fig:climppi}
    \end{subfigure}
    \begin{subfigure}[t]{0.48\textwidth}
        \centering
        \includegraphics[width=0.95\textwidth]{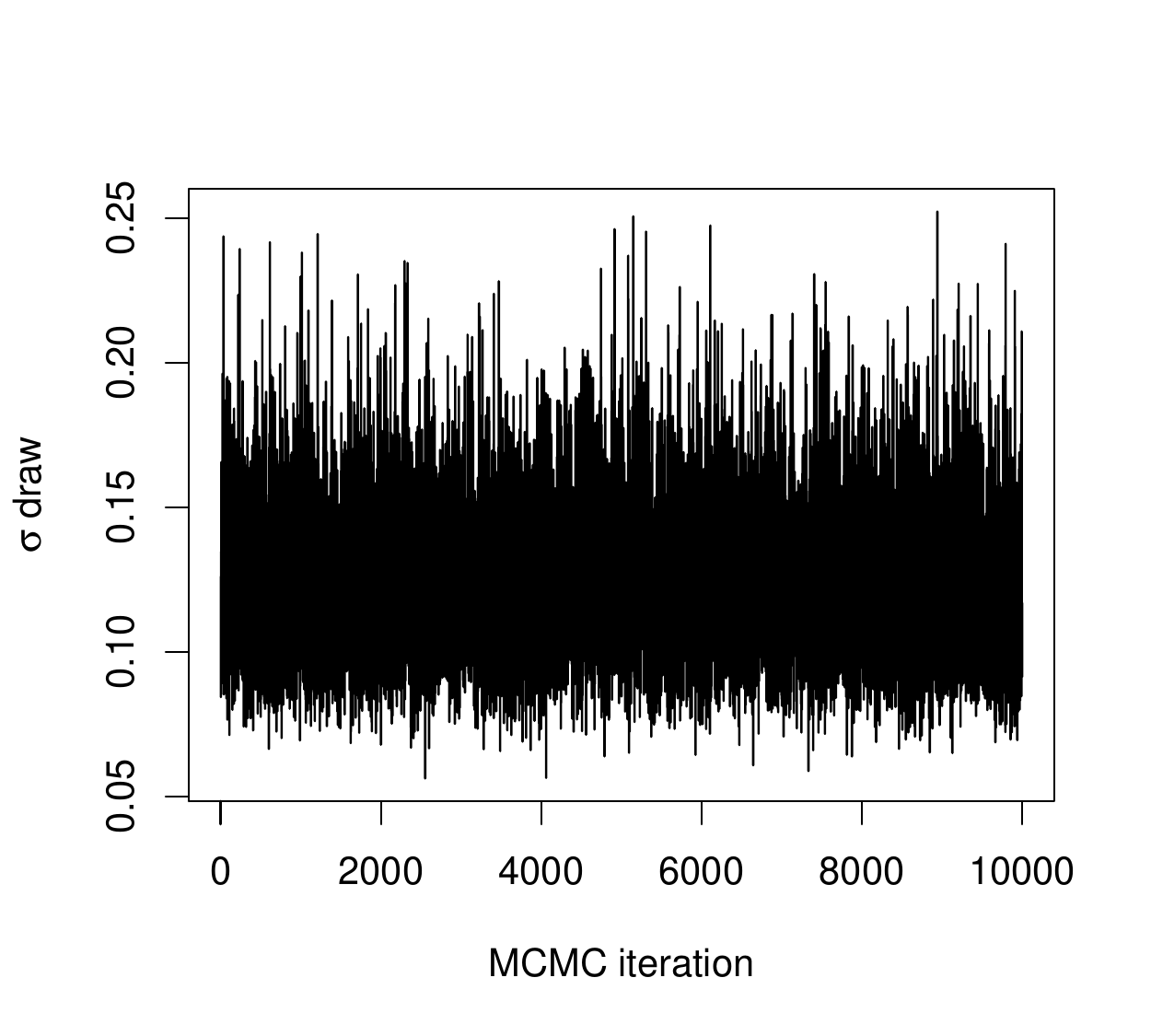}
        \caption{Trace plot for posterior draws of $\sigma$.}
        \label{fig:climsig}
    \end{subfigure}
    \caption{Diagnostic plots of BART model trained on data from En-ROADS climate simulator.}
    \label{fig:climdiagnostics}
\end{figure}

To assess prediction accuracy, we predicted temperature increases from an out-of-sample test set of 37 samples, which are chosen manually to achieve a wide range of true temperature increase values. The mean-squared prediction error of the mean BART and GP predictions at the 37 out-of-sample points are, respectively, $0.333$ and $0.968$. Figure \ref{fig:climppi} shows that the BART model accurately predicts temperature increase at, roughly, mid-range values of $f(\x) \in [2.5^{\circ} F, 7.5^{\circ} F]$. Outside of this range, the BART model tends to underpredict temperature increase. We suspect this underprediction issue at the upper range is due to the training points having a maximum global temperature increase value of $7.8^{\circ} F$ and hence can be fixed by adding more training samples with extreme response values, which can be done by using combinations of extreme values of the four most active input variables. The underprediction might also be fixed by increasing the prior variance of BART's terminal node parameters as discussed in \cite{Chipman12}. However, for this paper we use the default parameter settings in the \texttt{BART} R package \citep{BART19}.
We also see in Figure \ref{fig:climsig} that the $\sigma$ samples appear to be stationary, which implies MCMC convergence.

We conclude that in this no-noise application with $p = 11$ predictors, a sample size of $n = 10p$ suffices for a trained BART model to adequately capture input importance through its Sobol\'{} indices. Variable counts, on the other hand, do not provide enough evidence to convincingly order variables in terms of their importance.

%%%%%%%%%%%%%%%%%%%%%%%%%%%%%%%%%%%%%%%%%%%%%%%%%%%%%%%%%%%%%%%%%%%%%%%%%%%%%%%%%%%%%%%%%%%%%%%%%%%%%%%%%%%%%%%%%%%%%%%%%%%%%%%%%%%%%%%%%%%%%%%%%%%%%%%%%%%%%%%%%%%%%%%%%%%%%%%%%
\section{Summary and Discussion} 
\label{sec:discuss}

This paper has provided analytic expressions, explicit interpretations, and computational algorithms for determining Sobol\'{} indices for BART models. The indices are computed exactly and avoid Monte Carlo approximations. We showed the relationship between Sobol\'{} indices for BART models and sensitivity indices obtained from one-way counts, which are the predominant way of assessing input activity in BART (see \cite{Bleich14, Linero18}  among others). Theorem \ref{thm:mec} showed that under certain conditions, both the one-way count and the first-order Sobol\'{} index of variable $x_i$ are functions of the conditional expectation function $\E_{\X_{-i}}[\ens(\X; \{\T_t, \M_t\}_{t=1}^m) | X_i = \bm{\cdot}]$. We then quantify the properties of Sobol\'{} indices estimated from the BART model for five different analytic functions.
First the bias and the uncertainty of the BART-based Sobol\'{} indices
for estimating the true Sobol\'{} indices for the underlying $f(\bm{\cdot})$ are estimated. 
We find that the bias of the BART-based Sobol\'{} indices is largest with the multiplicative $g-$function than with any of the additive test functions. 
Then the rankings of variable activity as measured by the BART-based Sobol\'{} indices are compared with those provided by one-way counts. 
To make the second comparisons, we proposed a rank discrepancy $d_r$ to better suit the problem of comparing input activity assessments. 
We find that the first-order and total-effects BART-based Sobol\'{} indices empirically outperform one-way BART counts at capturing, respectively, a function's first-order and total-effects Sobol\'{} indices.

Finally, we applied our BART-based Sobol\'{} indices to data generated by the En-ROADS climate simulator to explore how to best reduce future global temperature increases. In particular we note that 32 of the 37 input values in Figure \ref{fig:climppi} and 109 of the 110 training inputs result in future global temperature increases above $1.5^{\circ} C$ ($2.7^{\circ} F$), which is the agreed upon upper limit of average global temperature increase above pre-industrial levels set by the 2016 Paris Agreement under the United Nations Framework Convention on Climate Change \citep{parisagreement}. In words, the vast majority of policy scenarios described in Section \ref{sec:app} will result in global temperature increases of at least $1.5^{\circ} C$ by the year 2100. Indeed, a 2018 report from the Intergovernmental Panel on Climate Change claims that this temperature increase will likely reach $1.5^{\circ} C$ between 2030 and 2052 if it increases at its current rate \citep{ipcc}. The IPCC report also details the global impact of a $1.5^{\circ} C$ increase. Given these drastic predictions, it is imperative to identify the most impactful factors in minimizing this temperature increase. To achieve a temperature increase below $1.5^{\circ} C$ by 2100, Figures \ref{fig:climsobol} and \ref{fig:climmaineffects} suggest maximizing carbon price and the energy efficiency of buildings and industry while minimizing economic growth and the use of methane and other gases (which includes nitrous oxide and fluorinated gases). 

This research suggests additional statistical investigations. 
\cite{Linero18} shows empirically that when a Dirichlet prior is used to generate variable selection probabilities for tree nodes, the posterior probability of an arbitrary inert variable being included in a BART model can drastically shrink. If inert variables are simply not used in a BART model's split rules, then Theorem \ref{thm:rmterms} tells us that computing all Sobol\'{} indices up to some order will require fewer calculations. Furthermore, excluding inert variables might also improve the accuracy or efficiency of our BART-based Sobol\'{} indices. These observations suggest 
comparing the accuracy of our BART-based Sobol\'{} indices using a Dirichlet prior with those obtained from the default prior as well as the effect of increasing sample sizes.  

As has been noted, \cite{Bleich14} and \cite{Linero18}, among others, 
use variable counts in their variable selection methods for BART models. We have seen in the En-ROADS climate simulator example that a trained BART model better captures input activity through its Sobol\'{} indices rather 
than through one-way counts. This example suggests that additional research is needed to study the specificity and sensitivity in selecting active inputs for the two methods in order definitively draw this conclusion.

We conclude with two important data/model extensions. Our derivation of the Sobol\'{} index calculations assumed that the input variables are uncorrelated. This is not true in many applications. For dependent input variables it will be very useful to derive analytic expressions for BART-based Sobol\'{} indices. Here  ideas from \cite{Kucherenko2012}, \cite{Mara2012}, and \cite{Glen2012}, who all discuss various ways to estimate 
Sobol\'{} sensitivity indices in the dependent input variable case will be of use. Finally, we note that most of the results in the paper should
extend to other tree ensemble methods, such as the random forest method described in \cite{Breiman01}.

Implementation of BART-based first-order, second-order, and total-effects Sobol\'{} indices can be found in the Open Bayesian Trees (OpenBT) project at \url{https://bitbucket.org/mpratola/openbt/} \citep{openbt}.

%%%%%%%%%%%%%%%%%%%%%%%%%%%%%%%%%%%%%%%%%%%%%%%%%%%%%%%%%%%%%%%%%%%%%%%%%%%%%%%%%%%%%%%%%%%%%%%%%%%%%%%%%%%%%%%%%%%%%%%%%%%%%%%%%%%%%%%%%%%%%%%%%%%%%%%%%%%%%%%%%%%%%%%%%%%%%%%%%
% \vspace{-.4in}
\begin{center}
	\section*{ACKNOWLEDGMENTS}
\end{center}

% \vspace{-.1in}

% A.H.~would like to acknowledge the Graduate School at The Ohio State University for support during the dissertation year. 
% The work of M.T.P.~was supported in part by the National Science Foundation under Agreement DMS-1916231 and in part by the King Abdullah University of Science and Technology (KAUST) Office of Sponsored Research (OSR) under Award No. OSR-2018-CRG7-3800.3.
% T.J.S.~would like to thank the Isaac Newton Institute for Mathematical Sciences for support and hospitality during the programme on {\it Uncertainty Quantification} when work on this paper was undertaken.
% This research was also sponsored, in part, by the National Science Foundation under Agreements DMS-0806134 and DMS-1310294 (The Ohio State University). 
We thank Professor Joseph Verducci for introducing A.H. to ranking models. We also thank the two referees and an Associate Editor for their comments, which have improved this paper.

Funding: This work was supported by the Graduate School at The Ohio State University; the National Science Foundation [Agreements DMS-1916231, DMS-0806134, DMS-1310294]; the King Abdullah University of Science and Technology (KAUST) Office of Sponsored Research (OSR) [Award No. OSR-2018-CRG7-3800.3]; and the Isaac Newton Institute for Mathematical Sciences.

% Funding: This work was supported by the National Institutes of Health [grant numbers xxxx, yyyy]; the Bill & Melinda Gates Foundation, Seattle, WA [grant number zzzz]; and the United States Institutes of Peace [grant number aaaa].

%%%%%%%%%%%%%%%%%%%%%%%%%%%%%%%%%%%%%%%%%%%%%%%%%%%%%%%%%%%%%%%%%%%%%%%%%%%%%%%%%%%%%%%%%%%%%%%%%%%%%%%%%%%%%%%%%%%%%%%%%%%%%%%%%%%%%%%%%%%%%%%%%%%%%%%%%%%%%%%%%%%%%%%%%%%%%%%%%
%% The Appendices part is started with the command \appendix;
%% appendix sections are then done as normal sections
\appendix
\section{Proofs of Theorems}

\begin{pfsobolbart}

According to Equation \ref{eq:vardecompdef}, the first step to computing the Sobol\'{} index for tree function $g(\bm{\cdot}; \T, \M)$ and variable index set $P \subset \{1, 2, \ldots, p\}$ is to compute the conditional expectation $\E_{\X_{-P}}[g(\X; \T, \M) \mid \X_P]$. By taking the appropriate conditional expectation of both sides of Equation \ref{eq:treebasis}, we get 
\begin{align}
\E_{\X_{-P}}[g(\X; \T, \M) \mid \X_P] 
= \sum_{k=1}^{|\M|} d_k^{-P} \1_{\R_k^P} (\X_P) 
\label{eq:treehfun},
\end{align} 
where hyperrectangle $\R_k^P = \prod_{i \in P} I_k^i$, and coefficients $d_k^{-P} = \mu_k \Prob_{-P}(\R_k^{-P})$. Due to Assumption \ref{cond:uncorr}, the coefficient expression simplifies to
\begin{equation}
d_k^{-P} = \mu_k \prod_{j \notin P} \Prob_j(I_k^j).
\label{eq:coeff}
\end{equation}

According to Equation \ref{eq:vardecompdef}, the first step to computing the Sobol\'{} index for ensemble function $\ens(\bm{\cdot}; \{\T_t, \M_t\}_{t=1}^m)$ and variable index set $P \subset \{1, 2, \ldots, p\}$ is to compute the conditional expectation $\E_{\X_{-P}}[\ens(\X; \{\T_t, \M_t\}_{t=1}^m) \mid \X_P]$. By taking the appropriate conditional expectation of both sides of Equation \ref{eq:enssot}, using linearity of expectations, and plugging in Equation \ref{eq:treehfun}, we get 
\begin{align*}
\E_{\X_{-P}}[\ens(\X; \{\T_t, \M_t\}_{t=1}^m) \mid \X_P]  
&= \sum_{t=1}^m \E_{\X_{-P}}[g(\X; \T_t, \M_t) \mid \X_P] \\
&= \sum_{t=1}^m \sum_{k=1}^{|\M_t|} d_{tk}^P \1_{\R_{tk}^P} (\X_P).
\end{align*}

It is more convenient to view this conditional expectation as a single sum over the ensemble's terminal nodes rather than as a double sum as shown above. Hence, we can express this conditional expectation as
\begin{equation}
\E_{\X_{-P}}[\ens(\X; \{\T_t, \M_t\}_{t=1}^m) \mid \X_P] = \sum_{k \in B_{\ens}} d_k^{-P} \1_{\R_k^P} (\X_P),
\label{eq:enshfun}
\end{equation} 
where $B_{\ens} = \cup_{t=1}^m B_{\T_t}$ is the index set over the terminal nodes of the trees in ensemble $\ens$.

Finally we are able to compute the variance terms in Equation \ref{eq:vardecompdef} for general variable index set $P \subset \{1, 2, \ldots, p\}$. First, we compute $\Var_{\X_P}\bigg( \E_{\X_{-P}}[\ens(\X; \{\T_t, \M_t\}_{t=1}^m) \mid \X_P] \bigg)$. Into this term we plug in Equation \ref{eq:enshfun}, apply the general result $\Var(U) = \Cov(U, U)$ for generic random variable $U$, and use the bilinearity property of covariance to get 
\begin{align*}
\Var_{\X_P}\bigg( \E_{\X_{-P}}[\ens(\X; \{\T_t, \M_t\}_{t=1}^m) \mid \X_P] \bigg)
&= \sum_{k \in B_{\ens}} \sum_{l \in B_{\ens}} d_k^{-P} d_l^{-P} \Cov_{\X_P}\Bigg(\1_{\R_k^P}(\X_P), \1_{\R_l^P}(\X_P)\Bigg),
\end{align*}
where the coefficients $d_k^{-P}$ and $d_l^{-P}$ are defined in Equation \ref{eq:coeff}. To each covariance term, which we will denote as $C^P_{k, l}$, we can apply the elementary covariance result $\Cov(U, V) = \E UV - \E U \E V$ for generic random variables $U$ and $V$ to get
\begin{equation*}
C^P_{k, l} = \Prob_P(\R_k^P \cap \R_l^P) - \Prob_P(\R_k^P) \Prob_P(\R_l^P).
% \label{eq:Sobolcovariance}
\end{equation*}
Thus, 
\begin{equation*}
\Var_{\X_P}\bigg( \E_{\X_{-P}}[\ens(\X; \{\T_t, \M_t\}_{t=1}^m) \mid \X_P] \bigg) = \sum_{k \in B_{\ens}} \sum_{l \in B_{\ens}} d_k^{-P} d_l^{-P} C^P_{k, l}.
% \label{eq:Sobolkernel}
\end{equation*}

In particular, when $P = \{i\}$, then 
\begin{align*}
\E_{\X_{-i}}[\ens(\X; \{\T_t, \M_t\}_{t=1}^m) \mid X_i] = \sum_{k \in B_{\ens}} d_k^{-i} \1_{I_k^i} (X_i), 
% \label{eq:ceoneway}
\end{align*}
where $d_k^{-i} = \mu_k \prod_{j \neq i} \Prob_j(I_k^j)$. 
The first-order Sobol\'{} index in Equation \ref{eq:vardecompdef} then becomes
\begin{equation*}
V_i = \sum_{k \in B_{\ens}} \sum_{l \in B_{\ens}} d_k^{-i} d_l^{-i} C^i_{k, l},
% \label{eq:Sobolkernel1}
\end{equation*}
where
\begin{equation*}
C^i_{k, l} = \Prob_i(I_k^i \cap I_l^i) - \Prob_i(I_k^i) \Prob_i(I_l^i).
% \label{eq:Sobolcovariance1}
\end{equation*}

\end{pfsobolbart}

\begin{pfmec}
Consider a BART ensemble $\ens_0$ with $m$ regression trees, where each tree is simply a terminal node with one terminal node parameter. Thus, the ensemble $\ens_0$ predicts the same value for any input $\x \in D$ and is hence a constant-mean model. 
Then any BART ensemble $\ens$ with $m$ regression trees can be thought of as $\ens_0$ having undergone a sequence of birth processes. 
Any birth process slices a terminal node's corresponding hyperrectangle into two smaller hyperrectangles according to some split rule. If we call this split rule ``$x_i < c$'', then this slice occurs on the $(p-1)$-dimensional hyperplane $x_i = c$ in $D$. The resulting ``left'' hyperrectangle gains a terminal node parameter $\mu_{left}$ while the resulting ``right'' hyperrectangle gains a terminal node parameter $\mu_{right}$.
Thus, if prior to the birth process the piecewise-constant function $\E_{\X_{-i}}[\ens(\X; \{\T_t, \M_t\}_{t=1}^m) \mid X_i = \bm{\cdot}]$ is constant at $x_i = c$ (i.e. the split rule ``$x_i < c$'' does not already exist in the ensemble) and $\mu_{left} \neq \mu_{right}$ (which is true almost surely but is also ensured through assumption \ref{cond:as}), then the birth process produces a jump in the piecewise-constant function at $x_i = c$. Meanwhile, the birth process does not produce a jump in any of the other piecewise-constant functions $\E_{\X_{-j}}[\ens(\X; \{\T_t, \M_t\}_{t=1}^m) \mid X_j = \bm{\cdot}]$ (for $j \neq i$). Hence, under the mentioned conditions, each birth process that produces a unique split rule that involves variable $x_i$ increments the number of jumps in the piecewise-constant function $\E_{\X_{-i}}[\ens(\X; \{\T_t, \M_t\}_{t=1}^m) \mid X_i = \bm{\cdot}]$ by one. 
\end{pfmec}

\begin{pfmew}
We have the following transformations to the conditional expectation function $\E_{\X_{-i}}[\ens(\X; \{\T_t, \M_t\}_{t=1}^m) \mid X_i = \bm{\cdot}]$: 
\begin{enumerate}
    \item First, center and scale the $e_{k^*}^i$. 
    Let 
    \begin{equation*}
    \tilde{e^i}_{k^*} = \sqrt{|B_{\ens}^i|} \frac{e_{k^*}^i - \bar{e_{\bm{\cdot}}^i}}{s}, 
    \end{equation*}
    where $\bar{e_{\bm{\cdot}}^i} = |B_{\ens}^i|^{-1} \sum_{k^* \in B_{\ens}^i} e_{k^*}^i$ and 
    $s^2 = [\sum_{k^* \in B_{\ens}^i} (e_{k^*}^i - \bar{e_{\bm{\cdot}}^i})^2] / (|B_{\ens}^i|-1)$ is the corrected sample variance of the $\{e_{k^*}^i\}$. Note that for any two indices $k^*, l^* \in B_{\ens}^i$, the relation $\tilde{e^i}_{k^*} = \tilde{e^i}_{l^*}$ holds if and only if $e^i_{k^*} = e^i_{l^*}$.
    \item Second, assign equal probability mass $|B_{\ens}^i|^{-1}$ to each $\I_{k^*}^i$. Introduce new intervals $\tilde{\I^i}_{k^*}$ by shifting and scaling $\I_{k^*}^i$ so that 
    \begin{enumerate}
        \item $\{\tilde{\I^i}_{k^*}\}_{k^* \in B_{\ens}^i}$ still partitions $I_D^i$ into exactly $|B_{\ens}^i|$ sets, and
        \item $\Prob_i(\tilde{\I^i}_{k^*}) = |B_{\ens}^i|^{-1}$ for all $k^* \in B_{\ens}^i$.
    \end{enumerate}
\end{enumerate}
Now define $\tilde{h_{\ens}^i}(X_i) := \tilde{\E}_{\X_{-i}}[\ens(\X; \{\T_t, \M_t\}_{t=1}^m) \mid X_i] = \sum_{k^* \in B_{\ens}^i} \tilde{e^i}_{k^*} \1_{\tilde{\I^i}_{k^*}}(\bm{X_i})$.
Using previous definitions, we have 
\begin{align*}
\Var_{X_i} \big( \tilde{h_{\ens}^i}(X_i) \big) 
&= \Var_{X_i} \bigg( \sum_{k^* \in B_{\ens}^i} \tilde{e^i}_{k^*} \1_{\tilde{\I^i}_{k^*}}(X_i) \bigg) \\
&= \sum_{k^* \in B_{\ens}^i} \sum_{l^* \in B_{\ens}^i} \tilde{e^i}_{k^*} \tilde{e^i}_{l^*} 
\Bigg[\Prob_i(\tilde{\I^i}_{k^*} \cap \tilde{\I^i}_{l^*}) 
- \Prob_i(\tilde{\I^i}_{k^*}) \Prob_i(\tilde{\I^i}_{l^*}) 
\Bigg].
\end{align*}

Recall that the intervals $\tilde{\I^i}_{k^*}$ still partition the original domain $I_D^i$. So if $k^* \neq l^*$, then $\tilde{\I^i}_{k^*} \cap \tilde{\I^i}_{l^*} = \emptyset$ and hence $\Prob_i(\tilde{\I^i}_{k^*} \cap \tilde{\I^i}_{l^*}) = 0$. Thus, 
\begin{align*}
\sum_{k^* \in B_{\ens}^i} \sum_{l^* \in B_{\ens}^i} \tilde{e^i}_{k^*} \tilde{e^i}_{l^*}
\Prob_i(\tilde{\I^i}_{k^*} \cap \tilde{\I^i}_{l^*}) 
&= \sum_{k^* \in B_{\ens}^i} (\tilde{e^i}_{k^*})^2 \Prob_i(\tilde{\I^i}_{k^*}).
\end{align*}

Since each interval $\tilde{\I^i}_{k^*}$ has equal probability mass, each $\Prob_i(\tilde{\I^i}_{k^*})$ becomes simply $|B_{\ens}^i|^{-1}$. So then 
\begin{align*}
\Var_{X_i} \big( \tilde{h_{\ens}^i}(X_i) \big)
&= |B_{\ens}^i|^{-1} \sum_{k^* \in B_{\ens}^i} (\tilde{e^i}_{k^*})^2 - 
|B_{\ens}^i|^{-2} \sum_{k^* \in B_{\ens}^i} \sum_{l^* \in B_{\ens}^i} \tilde{e^i}_{k^*} \tilde{e^i}_{l^*}.
\end{align*}

Note that the coefficients $\tilde{e_{k^*}^i}$ are centered so that the sum $\sum_{k^* \in B_{\ens}^i} \tilde{e_{k^*}^i}$ (and hence the double sum term in the preceding equation) equals zero. Also note that the $\tilde{e_{k^*}^i}$ are scaled so that $\sum_{k^* \in B_{\ens}^i} (\tilde{e_{k^*}^i})^2 = |B_{\ens}^i| (|B_{\ens}^i| - 1)$. We then have
\begin{align*}
\Var_{X_i} \big( \tilde{h_{\ens}^i}(X_i) \big)
&= |B_{\ens}^i| - 1.
\end{align*}

Now let $C_{\ens}^i$ be the set of all unique cutpoints involved in any split rule in $\ens$ that includes variable $x_i$. 
Recall that the set $B_{\ens}^i$ indexes a set of intervals that partition the domain's $i$th margin, i.e. 
the set $B_{\ens}^i$ indexes the set of intervals $\I_{k^*}^i = [\gamma_1, \gamma_2)$ (or $[\gamma_1, \gamma_2]$ if $\gamma_2 = b_D^i$), where $\gamma_1$ and $\gamma_2$ are any two consecutive (in value) points in $C_{\ens}^i \cup \{a_D^i, b_D^i\}$.
Between any two such partitioning intervals must be a cutpoint in $C_{\ens}^i$. Furthermore, by the assumption that $e_{k^*}^i = e_{l^*}^i$ implies $\I_{k^*}^i = \I_{l^*}^i$ for any indices $k^*, l^* \in B_{\ens}^i$, all cutpoints in $C_{\ens}^i$ must lie between two such partitioning intervals. That is, no cutpoint in $C_{\ens}^i$ can lie in the interior of any such partitioning interval. Thus, $|B_{\ens}^i| - 1 = |C_{\ens}^i|$, which is simply the number of unique split rules in $\ens$ that include variable $x_i$ and hence, by Theorem \ref{thm:mec}, equals the number of jumps in the original conditional expectation function, which equals the number of jumps in the transformed conditional expectation function.
\end{pfmew}

\begin{pfrmterms}
For any terminal node $k \in B_{\ens}$ where $v(k) \cap P = \emptyset$, the random quantity $\E[\mu_{k} \1_{\R_k} (\X) \mid \X_P]$ is in fact constant almost surely. Therefore, 
\begin{align*}
&\Var_{\X_P}( \E[\ens(\X; \{\T_t, \M_t\}_{t=1}^m) - \mu_{k} \1_{\R_k} (\X) \mid \X_P] ) \\
&= \Var_{\X_P}( \E[\ens(\X; \{\T_t, \M_t\}_{t=1}^m) \mid \X_P] - \E[\mu_{k} \1_{\R_k} (\X) \mid \X_P]) \\
&= \Var_{\X_P}( \E[\ens(\X; \{\T_t, \M_t\}_{t=1}^m) \mid \X_P] ).
\end{align*}
The result of the theorem will follow by applying this argument to all such terminal nodes.
\end{pfrmterms}

\begin{pfdist}
In general, the Kemeny-Snell (KS) distance between rankings $\bm{\alpha} = (\alpha_1, \ldots, \alpha_p)$ and $\bm{\beta} = (\beta_1, \ldots, \beta_p)$ is defined to be 
\begin{equation*}
d_{KS}(\alpha, \beta) = \frac{1}{2} \sum_{i=1}^p \sum_{j=1}^p |A_{ij} - B_{ij}|
\end{equation*}
where 
\[ A_{ij} = \begin{cases} 
      1 & \text{$\bm{\alpha}$ prefers object $i$ to object $j$} \\
      -1 & \text{$\bm{\alpha}$ prefers object $j$ to object $i$} \\
      0 & \text{$\bm{\alpha}$ prefers objects $i$ and $j$ equally} \\
   \end{cases}
\]
and $B_{ij}$ is similarly defined for ranking $\bm{\beta}$. 

For the rest of the proof, we will assume that rankings $\bm{\alpha}$ and $\bm{\beta}$ each have no ties. We will also take ranking $\bm{\alpha}$ to be the reference vector and hence will, without loss of generality, assume $\bm{\alpha} = (1, 2, \ldots, p)$. We will also refer to the sum in Equation \ref{eq:vsum} as the discrepancy $d_r$. Finally, we will prove desired equality via induction.

We first note that these assumptions greatly simplify the KS distance. If we think of the values $A_{ij}$ (similarly $B_{ij}$) as constituting a $p \times p$ matrix $A$ (similarly $B$) whose $ij$ entry is $A_{ij}$ (similarly $B_{ij}$), then both matrices $A$ and $B$ are antisymmetric, which implies $|A_{ij} - B_{ij}| = |A_{ji} - B_{ji}|$ for all $i, j = 1, \ldots, p$ and $A_{ij} = B_{ij} = 0$ if $i = j$. 
Therefore, we may reformulate the KS distance as 
\begin{equation*}
d_{KS}(\bm{\alpha}, \bm{\beta}) = \sum_{i < j} |A_{ij} - B_{ij}|.
\end{equation*}

We now proceed with the proof by induction. Suppose $p=2$. Half the $KS$ distance is then $\frac{1}{2}|A_{12} - B_{12}|$, where $A_{12} = 1$, while the $AH$ distance becomes $W_1$ (since $W_p = 0$ by default). One of two cases may occur. If $\beta_1 < \beta_2$, then $\beta_1 = 1$ and $\beta_2 = 2$. In this case, both the $AH$ distance and half the $KS$ distance are zero. If $\beta_1 > \beta_2$, then $\beta_1 = 2$ and $\beta_2 = 1$. In this case, both the $AH$ distance and half the $KS$ distance are unity. We note that values $\beta_1$ and $\beta_2$ must be distinct due to ranking $\bm{\beta}$ having no ties. Thus, the induction hypothesis holds for $p = 2$. 

Now suppose the induction hypothesis holds for arbitrary $p-1 \geq 3$. The KS distance can be decomposed into 
\begin{equation*}
d_{KS}(\bm{\alpha}, \bm{\beta}) = d_{KS}(\bm{\alpha}_{-1}, \bm{\beta}_{-1}) + \sum_{j=2}^p |A_{1j} - B_{1j}|,
\end{equation*}
where we define $\bm{\alpha}_{-1} := (\alpha_2, \ldots, \alpha_p)$ and $\bm{\beta}_{-1}$ similarly for ranking $\bm{\beta}$. The the discrepancy $d_r$, due to its stagewise nature, can also be decomposed: 
\begin{equation*}
d_{AH}(\bm{\alpha}, \bm{\beta}) = W_1 + d_{AH}(\bm{\alpha}_{-1}, \bm{\beta}_{-1}),
\end{equation*}
where $W_1 = \beta_1 - 1$ by default. By assumption, half the KS distance between $\bm{\alpha}_{-1}$ and $\bm{\beta}_{-1}$ equals the discrepancy $d_r$ between the same two quantities. Hence, we need only prove that $\frac{1}{2} \sum_{j=2}^p |A_{1j} - B_{1j}| = \beta_1 - 1$ to complete the proof. 

First, we note that $A_{1j} = 1$ for all $j > 1$ and, since $B_{1j}$ is either $1$ or $-1$, the quantity $A_{1j} - B_{1j}$ is nonnegative. Thus, $|A_{1j} - B_{1j}| = 1 - B_{1j}$ for all $j > 1$. But $B_{1j}$ is simply $\1_{\beta_1 < \beta_j} - \1_{\beta_1 > \beta_j}$. Hence, $\sum_{j=2}^p B_{1j} = (p - \beta_1) - (\beta_1 - 1) = p - 2\beta_1 + 1$. Therefore, $\frac{1}{2} \sum_{j=2}^p |A_{1j} - B_{1j}| = \beta_1 - 1$.
\end{pfdist}

\begin{pfinert}
Suppose we have computed discordances $W_1, \ldots, W_{k-1}$ for some $k > q_0$ and wish to compute discordance $W_k$. Then the remaining considered items, each having input activity measure values of zero, all have ranking number 1 in ranking $\rho_f$. Since at least one remaining considered item has ranking number 1 in ranking $\rho_{\ens}$, we get $W_k = 1 - 1 = 0$. 
\end{pfinert}

\begin{pftied}
We will partition the discordances into three sets: $\{W_1, \ldots, W_{j-1}\}$, $\{W_j \ldots, W_{j+u-1}\}$, and $\{W_{j+u}, \ldots, W_{|\rho_f|}\}$. After letting, for all $k = 1, \ldots, j-1, j+u, \ldots, |\rho_f|$, item $i_k$ be the item removed from consideration after computing $W_k$ but (if $k < |\rho_f|$) before computing $W_{k+1}$, we will then prove the desired invariance to permutation $\phi$ for the three sets of discordances. 

First, consider discordances $W_1, \ldots, W_{j-1}$. These discordances depend only on the ranking numbers of items $i_1, \ldots, i_{j-1}$ in $\rho_f$ and in $\rho_{\ens}$. Because these ranking numbers are invariant to choice of permutation $\phi$, these discordances are also invariant to $\phi$. 

Now consider discordances $W_{j+u}, \ldots, W_{|\rho_f|}$. Similar to the previous set of discordances, these discordances depend only on the ranking numbers of items $i_{j+u}, \ldots, i_{|\rho_f|}$ in $\rho_f$ and in $\rho_{\ens}$. Because these ranking numbers are invariant to choice of permutation $\phi$, these discordances are also invariant to $\phi$. 

Finally, consider discordances $W_j \ldots, W_{j+u-1}$. Because items $i_{j+1}, \ldots, i_{j+u}$ (and no other items) have ranking number $j$ in $\rho_f$, these discordance values are $W_k = r_{(k)} - j + 1$ for $k = j, \ldots, j+u-1$, where $r_{(j)}, \ldots, r_{(j+u-1)}$ are the order statistics of ranking numbers $r_{j}, \ldots, r_{j+u-1}$. Because order statistics are invariant to permutations of the statistic values, these discordances are invariant to permutation $\phi$. 
\end{pftied}

\section{Figures for En-ROADS Climate Simulator}

\begin{figure}[t]
    \centering
    \begin{subfigure}[t]{0.3\textwidth}
        \centering
        \includegraphics[width=0.95\textwidth]{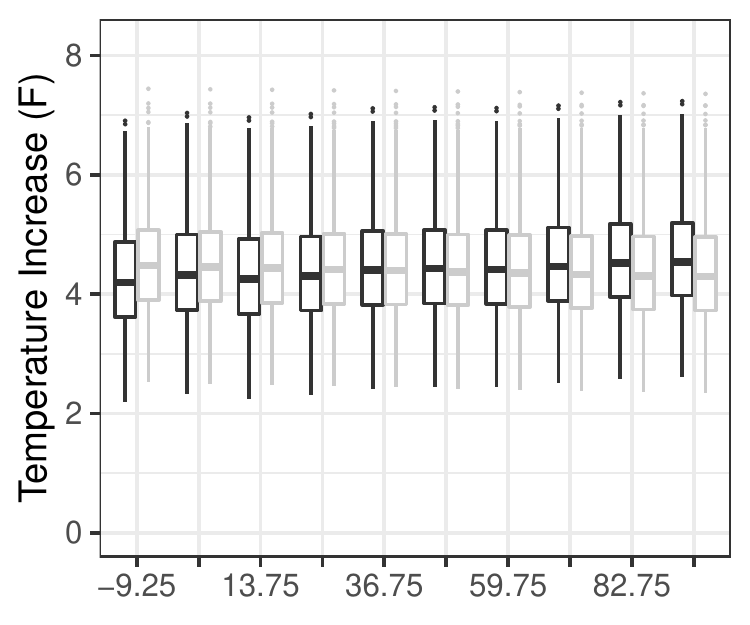}
        \caption{Oil.}
        % \label{fig:climrange1}
    \end{subfigure}
    \begin{subfigure}[t]{0.3\textwidth}
        \centering
        \includegraphics[width=0.95\textwidth]{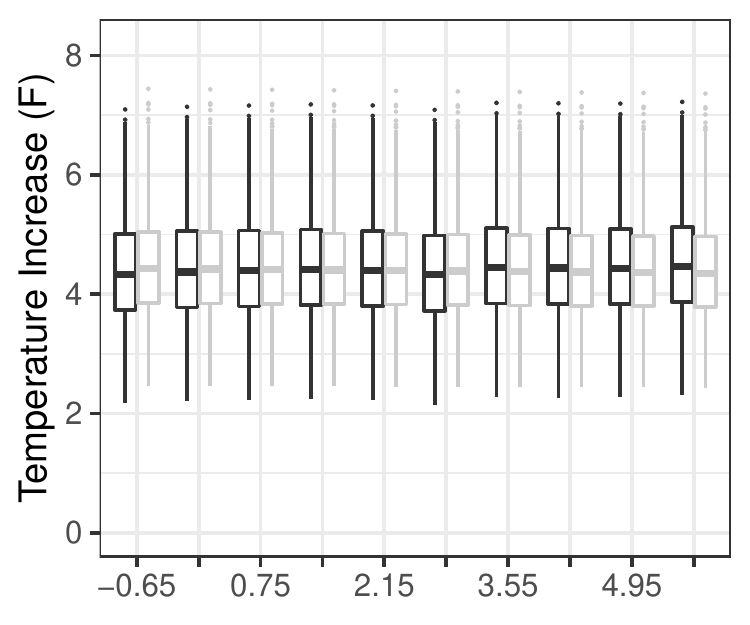}
        \caption{Gas.}
        % \label{fig:climrange2}
    \end{subfigure}
    \begin{subfigure}[t]{0.3\textwidth}
        \centering
        \includegraphics[width=0.95\textwidth]{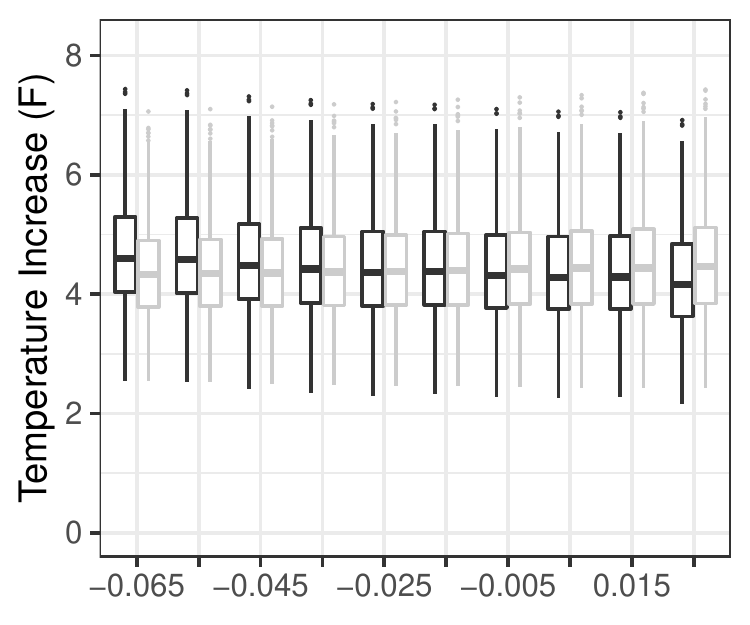}
        \caption{Renewables.}
        % \label{fig:climrange3}
    \end{subfigure}
    \begin{subfigure}[t]{0.3\textwidth}
        \centering
        \includegraphics[width=0.95\textwidth]{Plots/"03_range_Carbon_Price".pdf}
        \caption{Carbon Price.}
        % \label{fig:climrange4}
    \end{subfigure}
    \begin{subfigure}[t]{0.3\textwidth}
        \centering
        \includegraphics[width=0.95\textwidth]{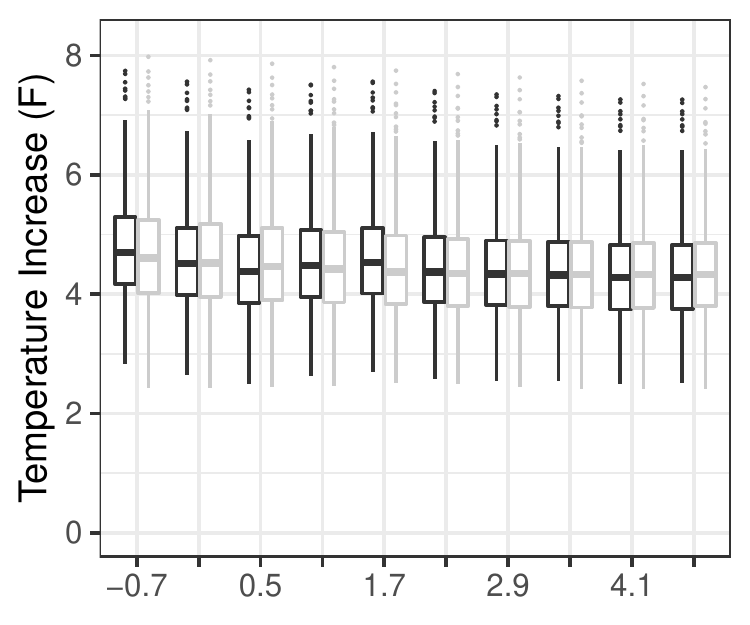}
        \caption{Energy Efficiency (Transport).}
        % \label{fig:climrange5}
    \end{subfigure}
    \begin{subfigure}[t]{0.3\textwidth}
        \centering
        \includegraphics[width=0.95\textwidth]{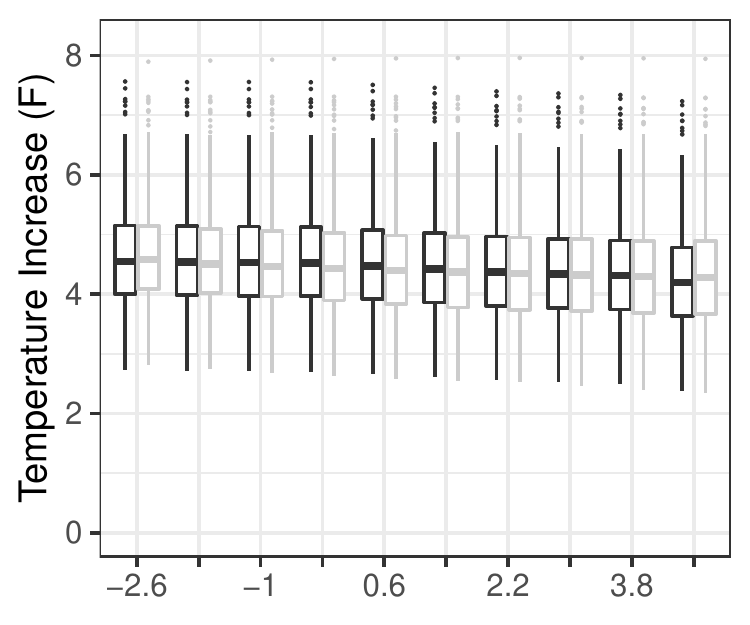}
        \caption{Electrification (Transport).}
        % \label{fig:climrange6}
    \end{subfigure}
    \begin{subfigure}[t]{0.3\textwidth}
        \centering
        \includegraphics[width=0.95\textwidth]{Plots/"04_range_En_Eff_Build".pdf}
        \caption{Energy Efficiency (Buildings and Industry).}
        % \label{fig:climrange7}
    \end{subfigure}
    \begin{subfigure}[t]{0.3\textwidth}
        \centering
        \includegraphics[width=0.95\textwidth]{Plots/"05_range_Econ_Growth".pdf}
        \caption{Economic Growth.}
        % \label{fig:climrange8}
    \end{subfigure}
    \begin{subfigure}[t]{0.3\textwidth}
        \centering
        \includegraphics[width=0.95\textwidth]{Plots/"06_range_Methane".pdf}
        \caption{Methane \& Other.}
        % \label{fig:climrange9}
    \end{subfigure}
    \begin{subfigure}[t]{0.3\textwidth}
        \centering
        \includegraphics[width=0.95\textwidth]{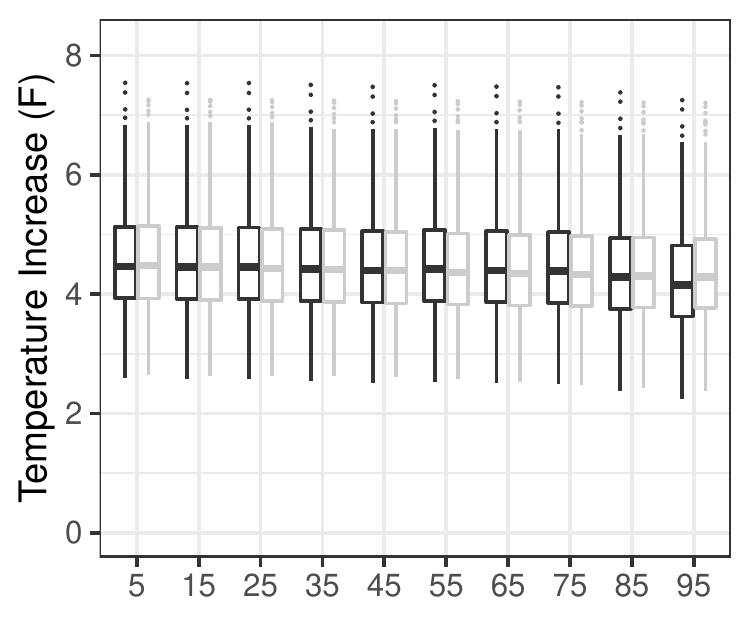}
        \caption{Afforestation.}
        % \label{fig:climrange10}
    \end{subfigure}
    \begin{subfigure}[t]{0.3\textwidth}
        \centering
        \includegraphics[width=0.95\textwidth]{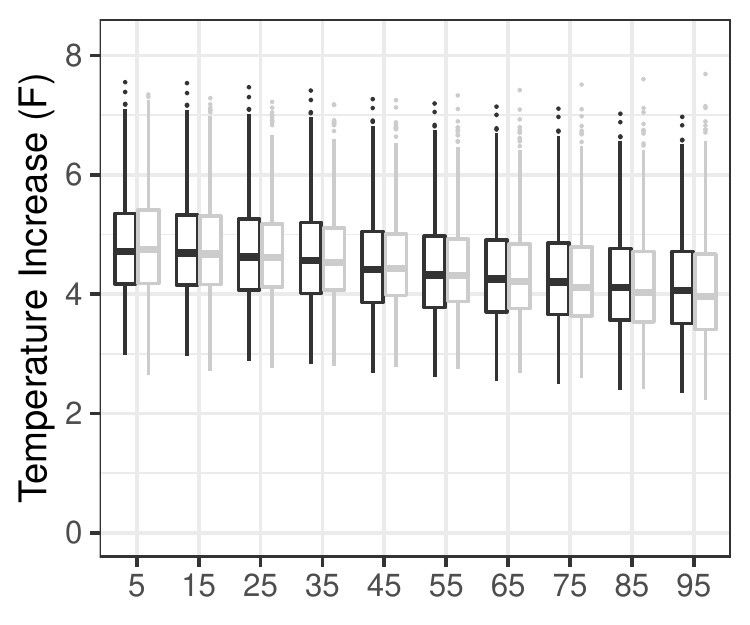}
        \caption{Technological (Carbon Removal).}
        % \label{fig:climrange11}
    \end{subfigure}
    \caption{Range plots for each of the 11 input variables.}
\end{figure}
% \XX{Describe mperk and how these plots were made.}

%% \section{}
%% \label{}

%% If you have bibdatabase file and want bibtex to generate the
%% bibitems, please use
%%
%%  \bibliographystyle{elsarticle-harv} 
%%  \bibliography{<your bibdatabase>}

%% else use the following coding to input the bibitems directly in the
%% TeX file.

\bibliographystyle{elsarticle-harv} 
\bibliography{bart} 

% \begin{thebibliography}{00}

% %% \bibitem[Author(year)]{label}
% %% Text of bibliographic item

% \bibitem[ ()]{}

% \end{thebibliography}
\end{document}